\documentclass[showpacs,preprintnumbers,amsmath,amssymb,twocolumn,aps,pre]{revtex4-1}
\usepackage{caption}
\usepackage{default}
\usepackage{amssymb}
\usepackage{color}
\usepackage{amsmath}
\usepackage{mathrsfs}
\usepackage{graphicx}
\usepackage{subfig}
\usepackage{epsfig}

\begin{document}
\title{Dark solitons in the Lugiato-Lefever equation with normal dispersion}
\author{P. Parra-Rivas$^{1,2}$, E. Knobloch$^3$, D. Gomila$^{2}$, and L. Gelens$^{1,4,5}$}

\affiliation{$^1$Applied Physics Research Group, APHY, Vrije Universiteit Brussel, 1050 Brussels, Belgium\\$^{2}$Instituto de F\'{\i}sica Interdisciplinar y Sistemas Complejos, IFISC (CSIC-UIB), Campus Universitat de les Illes Balears, E-07122 Palma de Mallorca, Spain\\$^3$Department of Physics, University of California, Berkeley CA 94720, USA\\$^4$Department of Chemical and Systems Biology, Stanford University School of 
Medicine, Stanford CA 94305, USA\\ $^5$Laboratory of Dynamics in Biological Systems, KU Leuven Department of Cellular and Molecular Medicine, University of Leuven, B-3000 Leuven, Belgium}

\date{\today}

\begin{abstract}
The regions of existence and stability of dark solitons in the Lugiato-Lefever model with normal chromatic dispersion are described. 
These localized states are shown to be organized in a bifurcation structure known as collapsed snaking implying the presence of
a region in parameter space with a finite multiplicity of dark solitons. For some parameter values dynamical instabilities are 
responsible for the appearance of oscillations and temporal chaos. The importance of the results for understanding frequency comb generation in microresonators is emphasized.
\end{abstract}

\maketitle

\section{Introduction}
Dark solitons, localized spots of lower intensity embedded in an homogeneous surrounding, are a particular type of solitons appearing in conservative or dissipative systems far from thermodynamic equilibrium \cite{Akhmediev}. In the latter case they are known as dissipative solitons (DSs) and related structures can be found in a large variety of systems, including those found in chemistry \cite{chemist}, gas discharges \cite{discharges}, fluid mechanics \cite{fluid}, vegetation and plant ecology \cite{vege}, as well as optics \cite{spatial_CS}, where they are known as cavity solitons. These structures arise as a result of a balance between nonlinearity and spatial coupling, and between driving and dissipation. In this work we focus on the field of optics, and study DSs in single mode fiber resonators and microresonators where they are known as temporal solitons \cite{leo_nat}. These systems are commonly described by the Lugiato-Lefever equation (LLE), a mean field model originaly introduced in \cite{lugiato_spatial_1987} 
in the context of ring cavities or a Fabry-Perot interferometer with transverse spatial extent, partially filled with a nonlinear medium. In temporal systems bright and dark solitons can be found. Taking into account only second order dispersion (SOD) two regimes can be identified, characterized by either normal or anomalous chromatic dispersion. In the latter case the only type of DSs that exist are bright solitons arising in both the monostable \cite{gomila_Scroggi} and bistable regimes \cite{leo-lendert,Parra_Rivas_2, Goday_chembo}. In contrast, in the normal SOD case the main type of DSs that appear are dark solitons \cite{Goday_chembo,Xue_NP,Lobanov_OE,KFC_dark}. In this work we provide a detailed analysis of the bifurcation structure and stability of dark DSs appearing in the normal dispersion regime, classifying the different dynamical regimes arising in this system. 

The organization of this paper is as follows. In Sec.~\ref{Sec::model}, we introduce the Lugiato-Lefever model in the context of temporal dynamics in fiber resonators and microresonators. We then analyze the spatial dynamics of spatially uniform states (Sec.~\ref{Sec::spady}), followed in Sec.~\ref{Sec::diagrams} by an analysis of the bifurcation structure of dark solitons. In Sec.~\ref{Sec::osci} we analyze oscillatory and chaotic dynamics of dark solitons. We conclude in Sec.~\ref{Sec::conclusions} by discussing the generality of the analysis provided in the earlier sections and in particular its relevance to frequency combs in nonlinear optics.

\section{The Lugiato-Lefever equation}\label{Sec::model}

\begin{figure}[tbp]
\centering
\includegraphics[width=8cm]{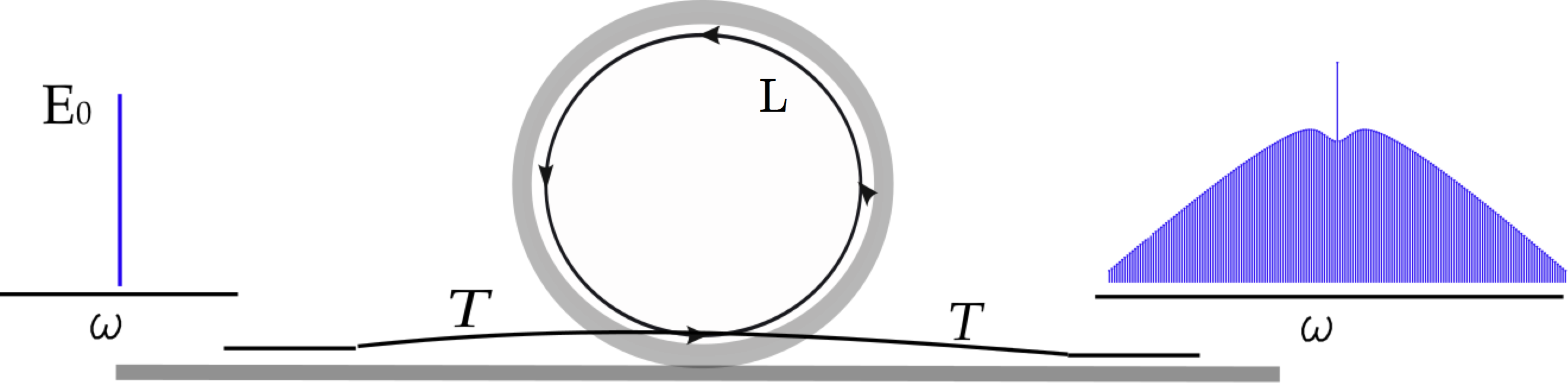}
\captionsetup{justification=raggedright,singlelinecheck=false}
\caption{(Color online) A synchronously pumped fiber cavity. Here $T$ is the transmission coefficient of the beam splitter and $L$  is the length of the fiber.}
\label{sketch}
\end{figure}

\begin{figure*}[t!]
\centering
\includegraphics[scale=0.9]{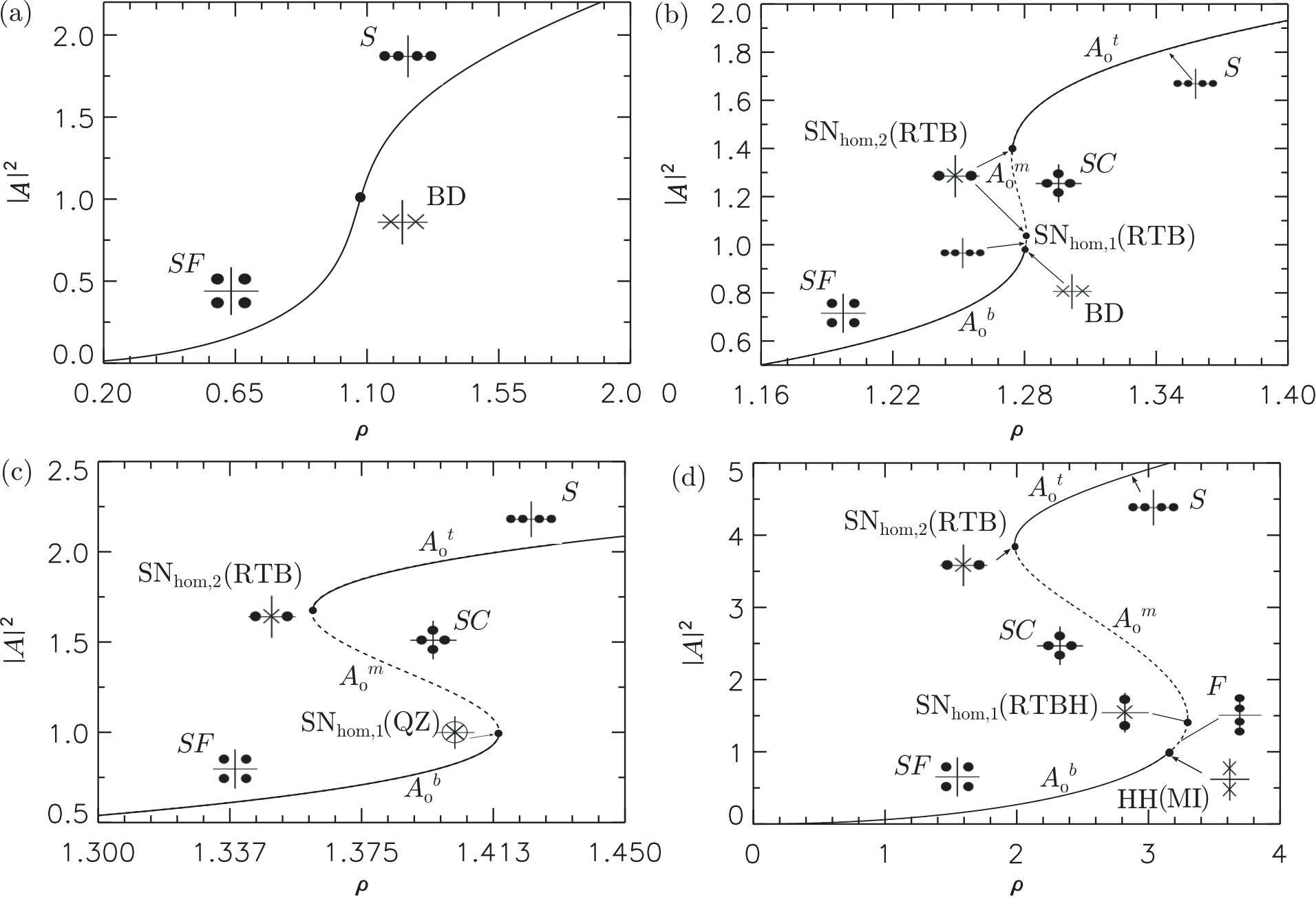}
\captionsetup{justification=raggedright,singlelinecheck=false}
\caption{Spatial eigenvalues of $A_0$ for several values of $\theta$. (a) $\theta=1.4<\sqrt{3}$; (b) $\sqrt{3}<\theta=1.8<2$; (c) $\theta=2$; (d) $2<\theta=4$. The different labels are explained in Table \ref{tab:template}. Solid (dashed) lines indicate stability (instability) in time.}
\label{spa_eigen_normal}
\end{figure*}

In this section we provide a brief introduction to the LLE in the context of fiber resonators and microresonators. We then employ the normalization of \cite{leo-lendert} to study the continuous wave (CW) or equivalently the homogeneous steady state (HSS) solutions of this model and to determine their temporal stability. Figure~\ref{sketch} shows a fiber cavity of length $L$ with a beam splitter with transmission coefficient $T$ and a continuous wave source of amplitude $E_0$.  At the beam splitter, the pump is coupled to the electromagnetic wave circulating inside the fiber. Under these conditions the evolution of the electric field $E\equiv E(t',\tau)$ within the cavity is described by the following evolution equation \cite{Haelterman}, 
\begin{equation}\label{model}
 t_R\frac{\partial E}{\partial t'}=-(\alpha+i\delta_0)E-i\frac{L\beta_2}{2}\frac{\partial^2E}{\partial\tau^2}
 +i\gamma L |E|^2E+\sqrt{T}E_0,
\end{equation}
where $\alpha>0$ describes the total cavity losses, $\beta_2$ is the second order dispersion coefficient ($\beta_2>0$ in the normal dispersion case while $\beta_2<0$ in the anomalous case), $\gamma>0$ is a nonlinear coefficient arising from the Kerr effect in the resonator, and $\delta_0$ is the cavity detuning. Here $\tau$ is the fast time describing the temporal structure of the nonlinear waves while the slow time $t'$ corresponds to the evolution time scale over many round-trips. After normalizing Eq.~(\ref{model}) we arrive to the dimensionless mean-field LLE \cite{lugiato_spatial_1987}:
\begin{equation}\label{LLE}
 \partial_tA=-(1+i\theta)A+i\nu\partial_x^2A+i|A|^2A+\rho,
\end{equation}
where $A(x,t)\equiv E(t',\tau)\sqrt{\gamma L/\alpha}$ is a complex scalar field, $t\equiv \alpha t'/t_R$, $x\equiv\tau\sqrt{2\alpha/(L|\beta_2|)}$, 
$\rho=E_0\sqrt{\gamma L T/\alpha^3}$, and $\theta=\delta_0/\alpha$. In the following we refer to the variable $x$ as a spatial coordinate
by analogy with other resonantly driven systems such as the LLE for spatially extended optical cavities \cite{lugiato_spatial_1987,spatial_CS} 
or the parametrically forced Ginzburg-Landau equation \cite{BuYoKn}.

Owing to the periodic nature of fiber cavities and microresonators, we consider periodic boundary conditions, i.e., $A(0,t)=A(L,t)$, 
where $L$ is now the dimensionless length of the system and choose $L=160$ for all numerical computations. The parameters 
$\rho,\theta\in\mathbb{R}$ correspond to the normalized injection and detuning, respectively, and serve as the control 
parameters of this system. The parameter $\nu$ represents the SOD coefficient and is also normalized: $\nu=-1$ in 
the normal dispersion case and $\nu=1$ in the anomalous dispersion case \cite{gomila_Scroggi,leo-lendert,Parra_Rivas_2,Goday_chembo}.
The present work is restricted to the case $\nu=-1$.

The steady states of Eq.~(\ref{LLE}) are solutions of the ordinary differential equation (ODE) 
\begin{equation}\label{LLEsteady}
i\nu \displaystyle\frac{d^2A}{dx^2}-(1+i\theta)A+i|A|^2A+\rho=0
\end{equation}
and are either spatially uniform states (HSSs) or spatially nonuniform states, consisting either of a periodic pattern 
(a spatially periodic state PS) or spatially localized states (DSs). In this section we focus on the HSSs, $A\equiv A_0$, 
leaving for subsequent sections the study of the other states. The $A_0$ states solve the classic cubic 
equation of dispersive optical bistability, namely
\begin{equation}\label{HSS}
 I_0^3-2\theta I_0^2+(1+\theta^2)I_0=\rho^2,
\end{equation}
where $I_0\equiv|A_0|^2$. For $\theta<\sqrt{3}$, Eq.~(\ref{HSS}) is single-valued and hence the system is monostable 
(see Fig.~\ref{spa_eigen_normal}(a)). For $\theta>\sqrt{3}$, Eq.~(\ref{HSS}) is triply-valued as shown 
in Figs.~\ref{spa_eigen_normal}(b)-(d). The transition between the three different solutions occurs via the two saddle nodes 
SN$_{hom,1}$ and SN$_{hom,2}$ located at 
\begin{equation}
 I_{\pm}\equiv|A_{\pm}|^2=\frac{2\theta}{3}\pm\frac{1}{3}\sqrt{\theta^2-3}.
\end{equation}
In the following we will denote the bottom solution branch (from $I_0=0$ to $I_-$) by $A_0^b$, the middle branch between $I_-$ and $I_+$ by $A_0^m$ and the top branch by $A_0^t$ ($I_0>I_{+}$). In terms of the real, $U\equiv Re[A]$, and imaginary, $V\equiv Im[A]$, parts the HSSs $A=A_0$ take the form
\begin{equation}
\left[\begin{array}{c}
U_0 \\ V_0\end{array}\right]=\left[\begin{array}{c}
\displaystyle\frac{\rho}{1+(I_0-\theta)^2} \\ \displaystyle\frac{(I_0-\theta)\rho}{1+(I_0-\theta)^2}\end{array}\right].
\end{equation}

We next analyze the linear stability of the HSSs to perturbations of the form
\begin{equation}
\left[\begin{array}{c}
U\\V
\end{array}\right]=\left[\begin{array}{c}
U_0\\V_0
\end{array}\right]+\epsilon\left[\begin{array}{c}
a\\b
\end{array}\right]e^{ikx+\Omega t}+c.c.,
\end{equation}
where $k$ represents the wave number of the perturbation. We find that the growth rate $\Omega(k)$ is given by 
\begin{equation}
 \Omega(k)=-1\pm\sqrt{4I_0\theta-3I_0^2-\theta^2+(4I_0-2\theta)\nu k^2- k^4}. 
\end{equation}
It follows that in the monostable regime the $A_0$ solution is always stable 
while for $\sqrt{3}<\theta<2$ the $A_0^b$ and $A_0^t$ states are stable and 
$A_0^m$ is unstable. These results are reflected in the diagrams shown in 
Figs.~\ref{spa_eigen_normal}(a) and (b). However, when $\theta>2$ the $A_0^b$ 
branch becomes unstable at a steady state bifurcation with $k\neq0$. This Turing 
or Modulational instability (MI) occurs at $I_0=1$ and generates a stationary 
periodic wavetrain with wave number $k_0=\sqrt{\nu(2-\theta)}$; $A_0^m$ remains 
unstable while $A_0^t$ is always stable. From a spatial dynamics point of 
view (Sec.~\ref{Sec::spady}) the MI bifurcation corresponds to a 
Hamiltonian-Hopf bifurcation in space (HH). No Hopf bifurcations in time of 
the HSSs are possible.

\section{Spatial dynamics}\label{Sec::spady}

In this section we investigate the conditions that are necessary for the presence of exponentially localized states that approach $A_0$ as $x\rightarrow\pm\infty$. To obtain these conditions we first rewrite Eq.~(\ref{LLEsteady}) as a dynamical system,
\begin{equation}\begin{array}{l}\label{SD}
d_xU=\tilde{U}\\
d_xV=\tilde{V}\\
d_x\tilde{U}=\nu\left[V+\theta U-UV^2-U^3\right]\\
d_x\tilde{V}=\nu\left[-U+\theta V-VU^2-V^3+\rho\right],
\end{array}\end{equation}
and employ the approach of spatial dynamics, i.e., we think of the solutions of 
Eq.~(\ref{SD}) as evolving in $x$, the rescaled fast time, from $x=-\infty$ to 
$x=\infty$ \cite{gomila_Scroggi,Champneys,Haragus,Gelens_NL,BuYoKn}. Thus DSs 
correspond to homoclinic orbits of Eq.~(\ref{SD}).

\begin{figure}[t!]
\centering
\includegraphics[scale=1]{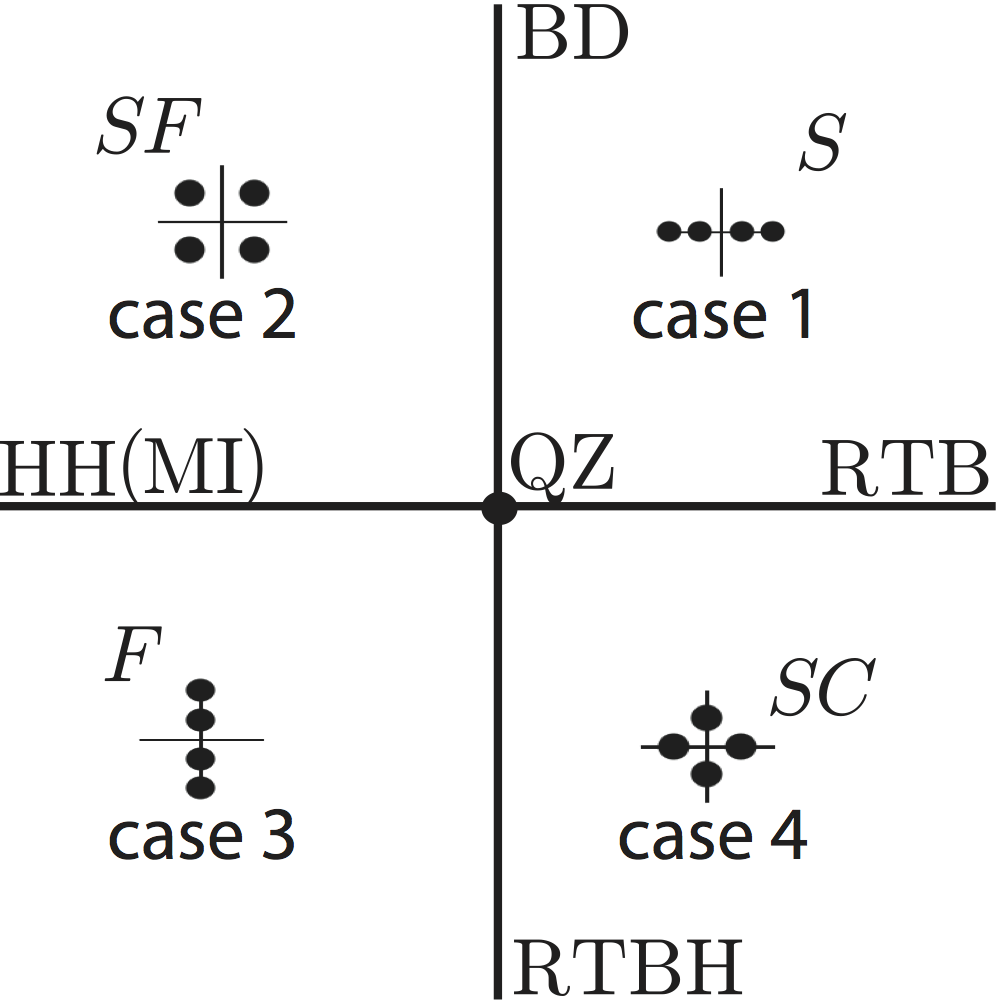}
\captionsetup{justification=raggedright,singlelinecheck=false}
\caption{Sketch of the possible organization of spatial eigenvalues $\lambda$ satisfying the biquadratic equation (\ref{biqua})
for a spatially reversible system. The names corresponding to the different labels are explained in Table \ref{tab:template}.}
\label{eigenvalue_regions}
\end{figure}

\begin{table}[t]
\centering
\begin{tabular}{|l|c|c|c|c|}
\hline
Cod & $(\lambda_{1,2,3,4})$  &  Name  & Label  \\
\hline
Zero & $(\pm q_0\pm i k_0)$  & Saddle-Focus & $SF$ \\
\hline
Zero & $(\pm q_1,\pm q_2)$  & Saddle & $S$ \\
\hline
Zero & $(\pm ik_1,\pm ik_2)$  & Center & $F$  \\
\hline
Zero & $(\pm q_0,\pm ik_0)$   & Saddle-Center & $SC$  \\
\hline
One & $(\pm q_0,0,0)$  & Rev.Takens-Bogdanov & RTB  \\
\hline
One & $(\pm ik_0,0,0)$ & Rev.Takens-Bogdanov-Hopf & RTBH   \\
\hline
One & $(\pm q_0,\pm q_0)$  &  Belyakov-Devaney & BD \\
\hline
One & $(\pm ik_0,\pm ik_0)$  & Hamiltonian-Hopf & HH(MI)\\
\hline
Two & $(0,0,0,0)$  & Quadruple Zero & QZ \\
\hline
\end{tabular}
\captionsetup{justification=raggedright,singlelinecheck=false}
\caption{Nomenclature used to refer to different transitions in the spatial eigenspectrum.}
\label{tab:template}
\end{table}

The fixed points of Eq.~(\ref{SD}) are the HSSs $A_0$ of the original evolution equation (\ref{LLE}). The stability of these 
fixed points (in space) is determined by the eigenspectrum of the Jacobian of the system (\ref{SD}) around $A_0\equiv U_0+iV_0$, 
namely 
\begin{equation}\label{SPL}
 \mathcal{J}=\nu\left[\begin{array}{cccc}
              0&0&\nu&0\\
               0&0&0&\nu\\
                \theta-V^2-3U^2&1-2UV&0&0\\
                 -(1+2UV)&\theta-U^2-3V^2&0&0\\
             \end{array}\right]_{(U_0,V_0)}.
\end{equation}
The four eigenvalues of $\mathcal{J}$ satisfy the biquadratic equation
\begin{equation}\label{biqua}
 \lambda^4+(4I_0-2\theta)\nu\lambda^2+\theta^2+3I_0^2-4\theta I_0+1=0.
\end{equation}
The form of this equation is a consequence of spatial reversibility \cite{Devaney,Homburg,Knobloch15}, i.e., the invariance of Eq.~(\ref{LLE}) under the transformation $(x,A)\mapsto(-x,A)$, or equivalently the invariance of the system (\ref{SD}) under $(x,U,V,\tilde{U},\tilde{V})\mapsto(-x,U,V,-\tilde{U},-\tilde{V})$. This invariance implies that if $\lambda$ is a spatial eigenvalue, so are $-\lambda$ and $\pm\lambda^*$, where $^*$ indicates complex conjugation. Consequently there are four possibilities:
\begin{enumerate}
 \item the eigenvalues are real: $\lambda_{1,2,3,4}=(\pm q_1,\pm q_2)$
 \item there is a quartet of complex eigenvalues: $\lambda_{1,2,3,4}=(\pm q_0 \pm ik_0)$
 \item the eigenvalues are imaginary: $\lambda_{1,2,3,4}=(\pm ik_1,\pm ik_2)$
 \item two eigenvalues are real and two imaginary: $\lambda_{1,2,3,4}=(\pm q_0,\pm ik_0)$.
\end{enumerate}

A sketch of these possible eigenvalue configurations is shown in Fig.\ \ref{eigenvalue_regions}, and their names and codimension are provided 
in Table \ref{tab:template}. The transition from case 1 to case 2 is through a Belyakov-Devaney (BD) \cite{Champneys,Haragus} point with 
eigenvalues $(\pm q_0,\pm q_0)$, while the transition from case 2 to case 3 is via a Hamiltonian-Hopf (HH) bifurcation \cite{Ioos,Haragus},
with $\lambda_{1,2,3,4}=(\pm ik_0,\pm ik_0)$. The transition from case 1 to case 4 is via a reversible Takens-Bogdanov (RTB) bifurcation with 
eigenvalues $\lambda_{1,2,3,4}=(\pm q_0,0,0)$ \cite{Champneys,Haragus} while the transition from case 3 to case 4 is via a reversible 
Takens-Bogdanov-Hopf (RTBH) bifurcation with eigenvalues $\lambda_{1,2,3,4}=(\pm ik_0,0,0)$ \cite{Champneys,Haragus}.
The unfolding of all these scenarios is related to the quadruple zero (QZ) codimension-2 point \cite{Champneys,Haragus}. 
As shown in the next section the transitions between these different regimes organize the 
parameter space for DSs.  

\begin{figure}[t!]
\centering
\includegraphics[scale=1]{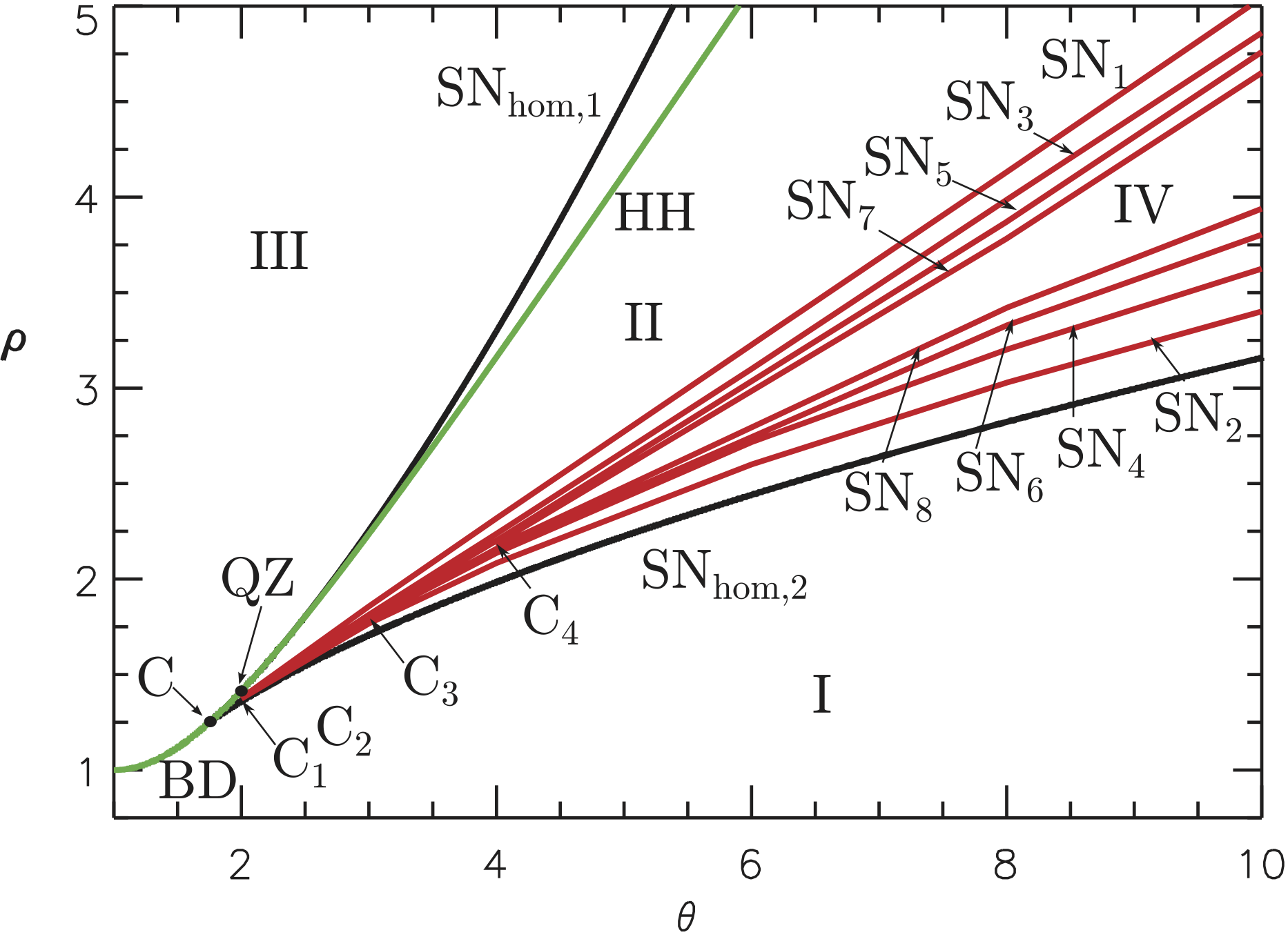}
\captionsetup{justification=raggedright,singlelinecheck=false}
\caption{(Color online) The $(\theta,\rho)$ parameter space for normal dispersion in the region of existence of dark solitons. The green line corresponds to the HH bifurcation, the black lines to SN bifurcations of the HSS, and the red lines to SN bifurcations of the dark DSs. The bifurcation lines and the regions I--IV are discussed in more detail in the text.}
\label{parameter_space}
\end{figure}

The eigenvalues satisfying Eq.~(\ref{biqua}) are
\begin{equation}
 \lambda=\pm\sqrt{(\theta-2I_0)\nu\pm\sqrt{I_0^2-1}}.
\end{equation}
Figure~\ref{spa_eigen_normal} summarizes the possible eigenvalue configurations for normal dispersion ($\nu=-1$). The transition at $I_0=1$, i.e., along the green curve
\begin{equation}
 \rho=\sqrt{1+(1-\theta)^2}
\end{equation}
in Fig.~\ref{parameter_space}, corresponds to a BD transition when $\theta<2$ and an HH transition when $\theta>2$. 
Figures~\ref{spa_eigen_normal}(a)-(b) correspond to the case $\theta<2$; we see that the saddle-node bifurcation at SN$_{hom,1}$
corresponds to a RTB bifurcation. In contrast, for $\theta>2$ SN$_{hom,1}$ has become a RTBH bifurcation 
(Fig.~\ref{spa_eigen_normal}(d)). For $\theta=2$ (Fig.~\ref{spa_eigen_normal}(c)) the BD, HH, RTB and RTBH lines meet at
the QZ point. In the parameter space of Fig.~\ref{parameter_space} the QZ point corresponds to $(\theta,\rho)=(2,\sqrt{2})$. The 
other relevant bifurcation lines in this scenario correspond to SN$_{hom,2}$. This point corresponds to a RTB bifurcation in 
space regardless of the value of $\theta$. 

In terms of spatial dynamics, DSs correspond to intersections of the stable and unstable manifolds of the HSS \cite{Knobloch15}. In 
cases 1 and 2 the HSS has a 2-dimensional stable and a 2-dimensional unstable manifold. These manifolds are transverse to the 2-dimensional 
fixed point subspace of the symmetry $(x,A)\mapsto(-x,A)$ and hence intersect in a structurally stable way. Therefore we expect DSs in cases 
1 and 2 only. In case 4, the stable and unstable manifolds of the HSS are 1-dimensional and DSs, although possible, are exceptional \cite{Kolossovski}. 
In Fig.~\ref{spa_eigen_normal}(b) DSs bifurcate from both SN$_{hom,1}$ and SN$_{hom,2}$. When HH is present (Fig.~\ref{spa_eigen_normal}(d))
DSs bifurcate from HH and from SN$_{hom,2}$. 

In Sec.~\ref{Sec::diagrams} we show that it is possible to compute DSs analytically near the bifurcation points that 
produce them, and use the resulting expressions to initialize numerical continuation \cite{algower} of these states.

\section{Bifurcations and existence of dissipative solitons}\label{Sec::diagrams}

\subsection{Weakly nonlinear analysis}

In this section we compute weakly nonlinear DSs using multiple scale perturbation theory near the RTB bifurcation corresponding to SN$_{hom,2}$. 
The procedure applies equally around the other RTB point at SN$_{hom,1}$. Following \cite{BuYoKn}, we fix the value of $\theta$ and suppose that
the DSs at $\rho\approx \rho_t$, where $\rho=\rho_t$ corresponds to the SN$_{hom,2}$ bifurcation, are captured by the ansatz $U=U^*+u$, $V=V^*+v$, 
where $U^*$ and $V^*$ represent the HSS $A_0^t$ and $u$ and $v$ capture the spatial dependence. We next introduce appropriate asymptotic expansions
for each variable in terms of a small parameter $\epsilon$ defined through the relation $\rho=\rho_t+\epsilon^2\delta$, where $\delta$ is defined in
the Appendix. Each variable is written in the form 
\begin{equation}
 \left[\begin{array}{c}
U^*\\V^*
\end{array}\right]= \left[\begin{array}{c}
U_t\\V_t
\end{array}\right]+ \epsilon\left[\begin{array}{c}
U_1\\V_1
\end{array}\right]+...
\end{equation}
and 
\begin{equation}
 \left[\begin{array}{c}
u\\v
\end{array}\right]= \epsilon\left[\begin{array}{c}
u_1\\v_1
\end{array}\right]+ \epsilon^2\left[\begin{array}{c}
u_2\\v_2
\end{array}\right]+...
\end{equation}
and these expressions inserted into Eq.~(\ref{LLEsteady}). Solving order by order in $\epsilon$ we find that the leading order asymptotic solution 
close to the RTB point is given by
\begin{equation}\label{asymp1}
 \left[\begin{array}{c}
U\\V
\end{array}\right]= \left[\begin{array}{c}
U_t\\V_t
\end{array}\right]+ \epsilon\left[\begin{array}{c}
U_1+u_1\\V_1+v_1
\end{array}\right],
\end{equation}
where $U_t$ and $V_t$ correspond to the HSS at $\rho = \rho_t$, and
\begin{equation}\label{asymp2}
\left[\begin{array}{c}
u_1\\v_1
\end{array}\right]= 
\left[\begin{array}{c}
U_1\\ V_1\end{array}\right]\psi(x),
\end{equation}
with
\begin{equation}\label{asymp3}
\left[\begin{array}{c}
U_1\\V_1
\end{array}\right]= 
\mu\left[\begin{array}{c}
1\\ \eta\end{array}\right]
\end{equation}
and 
\begin{equation}\label{sech}
\psi(x)= -3{\textnormal{sech}}^2\left[ \displaystyle\frac{1}{2}\sqrt{-\frac{\alpha_2}{\alpha_1}}
\left(\displaystyle\frac{\rho-\rho_{t}}{\delta}\right)^{1/4}x\right].
\end{equation}

\begin{figure}[t!]
\centering
\includegraphics[scale=1.0]{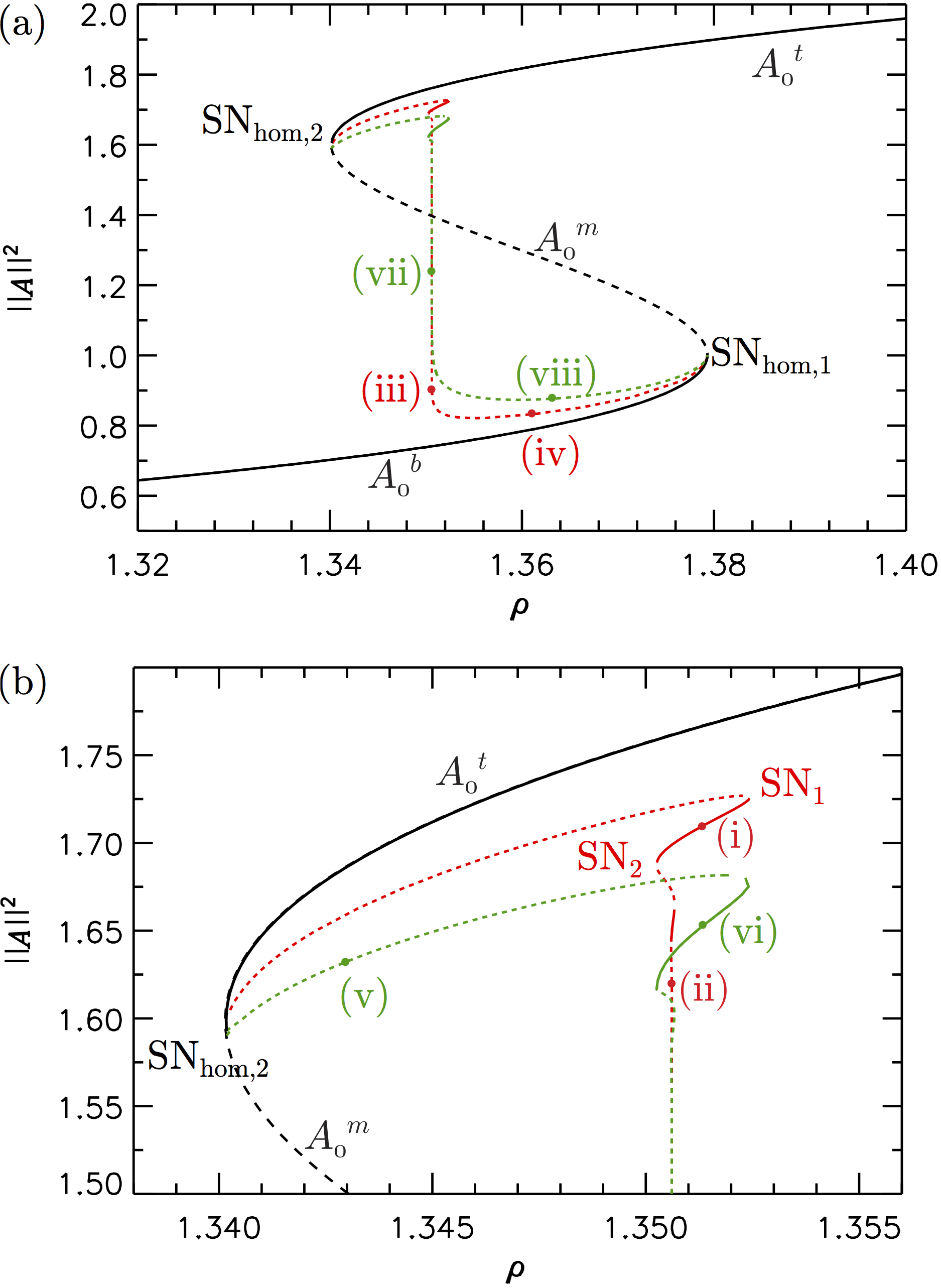}
\captionsetup{justification=raggedright,singlelinecheck=false}
\caption{(a) Bifurcation diagram at $\theta=1.95$. (b) Zoom of panel (a) around SN$_{hom,2}$. The homogeneous steady states
HSS are shown in black, 1-soliton states in red and 2-soliton states in green. Temporally stable (unstable) DSs are indicated 
using solid (dashed) lines. Profiles corresponding to the labeled locations are shown in Fig.~\ref{DarkS_1.95_profiles} and 
in more detail in Fig.~\ref{DarkS_1.95_profiles_detail}.}
\label{DarkS_1.95}
\end{figure}

Here $\eta$, $\mu$, $\alpha_1$ and $\alpha_2$ are parameters defined in the Appendix, where the details of the calculation 
can be found. The localized structure defined by the asymptotic solution is shown in Fig.~\ref{pulse_up} of the Appendix; 
the negative sign in Eq.~(\ref{sech}) implies that the solution is a hole in the background $A_0^t$ state, i.e., a dark 
soliton. Of course, on a large domain we expect to find states with 2 or more dark solitons as well. When these are well 
separated these states behave like 1-soliton states and so should bifurcate from the vicinity of SN$_{hom,2}$ just like the 
1-soliton states.

\begin{figure}[t!]
\centering
\includegraphics[scale=1.0]{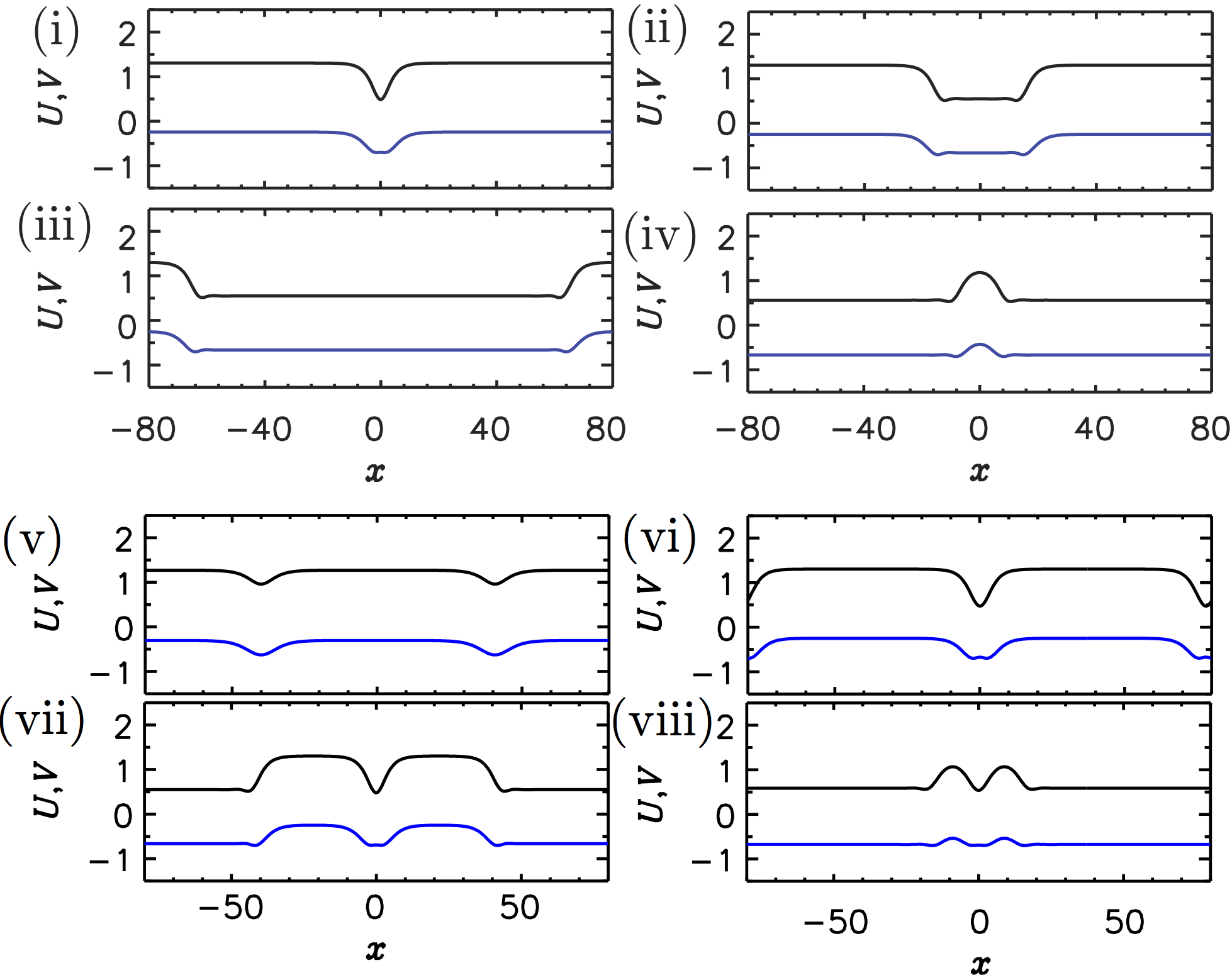}
\captionsetup{justification=raggedright,singlelinecheck=false}
\caption{(Color online) Spatial profiles of DSs (dark and bright 1-soliton and 2-soliton states) corresponding to the locations indicated 
in Fig.~\ref{DarkS_1.95}(a,b), with $U(x)$ in black and $V(x)$ in blue. Near SN$_{hom,2}$ the states resemble holes (dark solitons) 
while near SN$_{hom,1}$ they resemble localized pulses (bright solitons).}
\label{DarkS_1.95_profiles}
\end{figure}

\begin{figure*}[t!]
\centering
\includegraphics[scale=1.0]{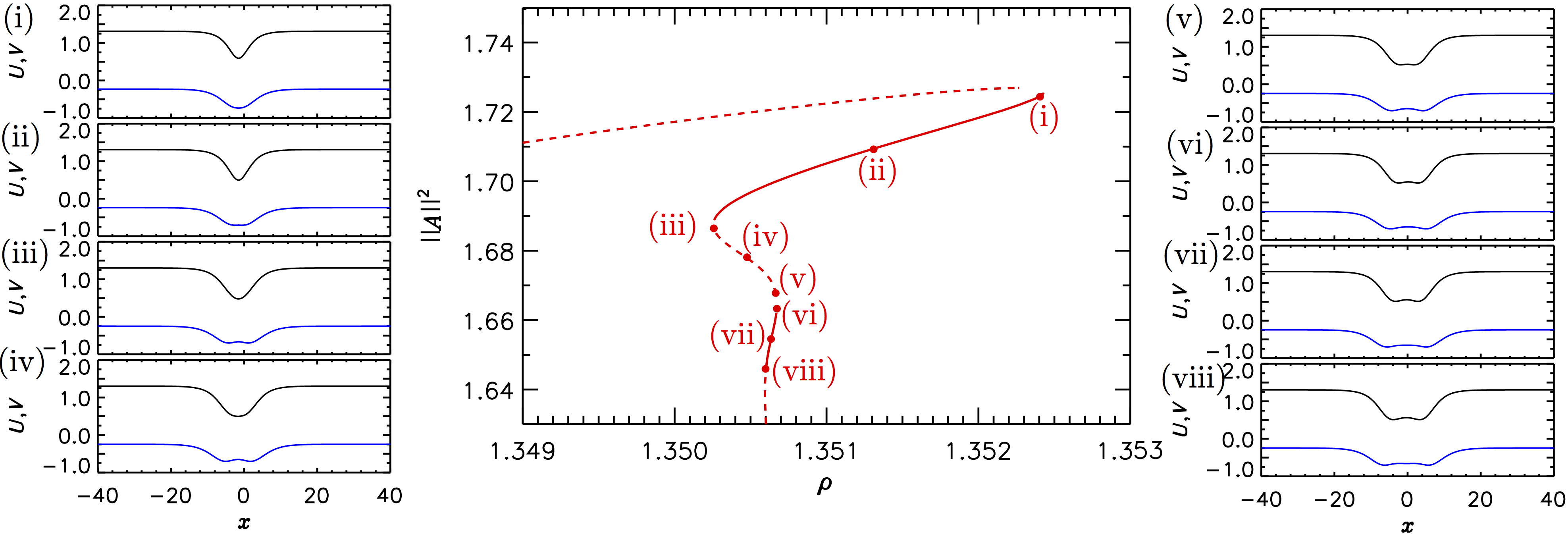}
\captionsetup{justification=raggedright,singlelinecheck=false}
\caption{(Color online) Spatial profiles of dark solitons near the upper end of the $\theta=1.95$ 1-soliton branch at locations indicated in the middle panel, showing that the splitting of the central peak (dip) in $(U(x),V(x))$, shown in black and red, respectively, occurs at different locations along the branch.}
\label{DarkS_1.95_profiles_detail}
\end{figure*}

We now discuss the bifurcation structure of dark solitons in two regimes: the bistable region before the
QZ point, namely for $\sqrt{3}<\theta<2$, and the bistable region after QZ, i.e., for $\theta>2$. 

\subsection{Dark solitons for $\sqrt{3}<\theta<2$}\label{Sec::bright}

In the following we use the $L^2$ norm, $||A||^2\equiv\frac{1}{L}\int_0^L |A|^2\, dx$, to represent the DSs in a bifurcation diagram. Figure~\ref{DarkS_1.95}(a), computed for $\theta=1.95$, reveals the presence of two soliton branches, the first of which consists of a single dark soliton in the domain (hereafter the 1-soliton state, red branch) while the second consists of a pair of identical equidistant dark solitons in the domain (hereafter the 2-soliton state, green branch). Figure~\ref{DarkS_1.95}(b) provides a zoom of the region near SN$_{hom,2}$. Both branches bifurcate from HSS very close to SN$_{hom,2}$ as anticipated in the preceding section and both undergo collapsed snaking \cite{KnWa,BuYoKn_colapsing}. Specifically, the 1-soliton state bifurcates from HSS closer to the saddle node at $\rho=\rho_t$ and its solution branch undergoes a series of exponentially decaying oscillations in the vicinity of a critical value of $\rho$, hereafter $\rho=\rho_M\approx 1.3506074$. During this process the hole corresponding to the dark soliton deepens, forming a pair of fronts connecting $A_0^t$ and $A_0^b$ and then broadens as the $A_0^b$ state expels $A_0^t$ (Fig.~\ref{DarkS_1.95_profiles}), becoming in an infinite system a heteroclinic cycle between $A_0^t$ and $A_0^b$ at $\rho_M$. In gradient systems this point corresponds to the so-called Maxwell point, where both homogeneous solutions have equal energy. In non-gradient systems, such as LLE, such a cycle may still be present, even though an energy cannot be defined, and we retain this terminology to refer to its location, i.e., the parameter value corresponding to the presence a pair of stationary, infinitely separated fronts connecting $A_0^t$ to $A_0^b$ and back again.

The successive saddle nodes seen in Fig.~\ref{DarkS_1.95} correspond to the appearance of additional oscillations in the tails of the fronts as the local maximum (minimum) at the symmetry point $x=0$ turns into a local minimum (maximum) and back again, and hence to a gradual increase in the width of the hole. Figure \ref{DarkS_1.95_profiles_detail} shows a detail of this process. The associated hole states are temporally stable between SN$_1$ and SN$_2$, and on all the subsequent branch segments with positive slope \cite{KnWa,BuYoKn_colapsing}, shown using solid lines. A profile of a stable localized hole on the SN$_1$--SN$_2$ segment is shown in Fig.~\ref{DarkS_1.95_profiles}(i). For the value of $\theta$ used in Fig.~\ref{DarkS_1.95} the collapse of the saddle nodes to $\rho_M$ is very abrupt because the spatial oscillations in the tail of the front decay very fast. Figure \ref{DarkS_4} shows a clearer example of the behavior in this region, albeit for a larger value of $\theta$. 

In finite systems the hole or 1-soliton branch departs from $\rho\approx \rho_M$ when the maximum amplitude starts to decrease below $A_0^t$ and the solution turns into a bright soliton sitting on top of $A_0^b$ (Fig.~\ref{DarkS_1.95_profiles}(iv)). The branch then terminates at SN$_{hom,1}$, where the amplitude of this soliton falls to zero. On an infinite domain the DS branches bifurcating from SN$_{hom,2}$ and SN$_{hom,1}$ remain distinct and do not connect up. Figures~\ref{DarkS_1.95} and \ref{DarkS_1.95_profiles} show that the 2-soliton branch behaves in a very similar manner. In fact, this is so for all $n$-soliton branches ($n\ge3$), provided the solitons remain sufficiently well separated.

\subsection{Dark solitons for $\theta>2$}\label{Sec::dark}

\begin{figure}[t!]
\centering
\includegraphics[scale=1.0]{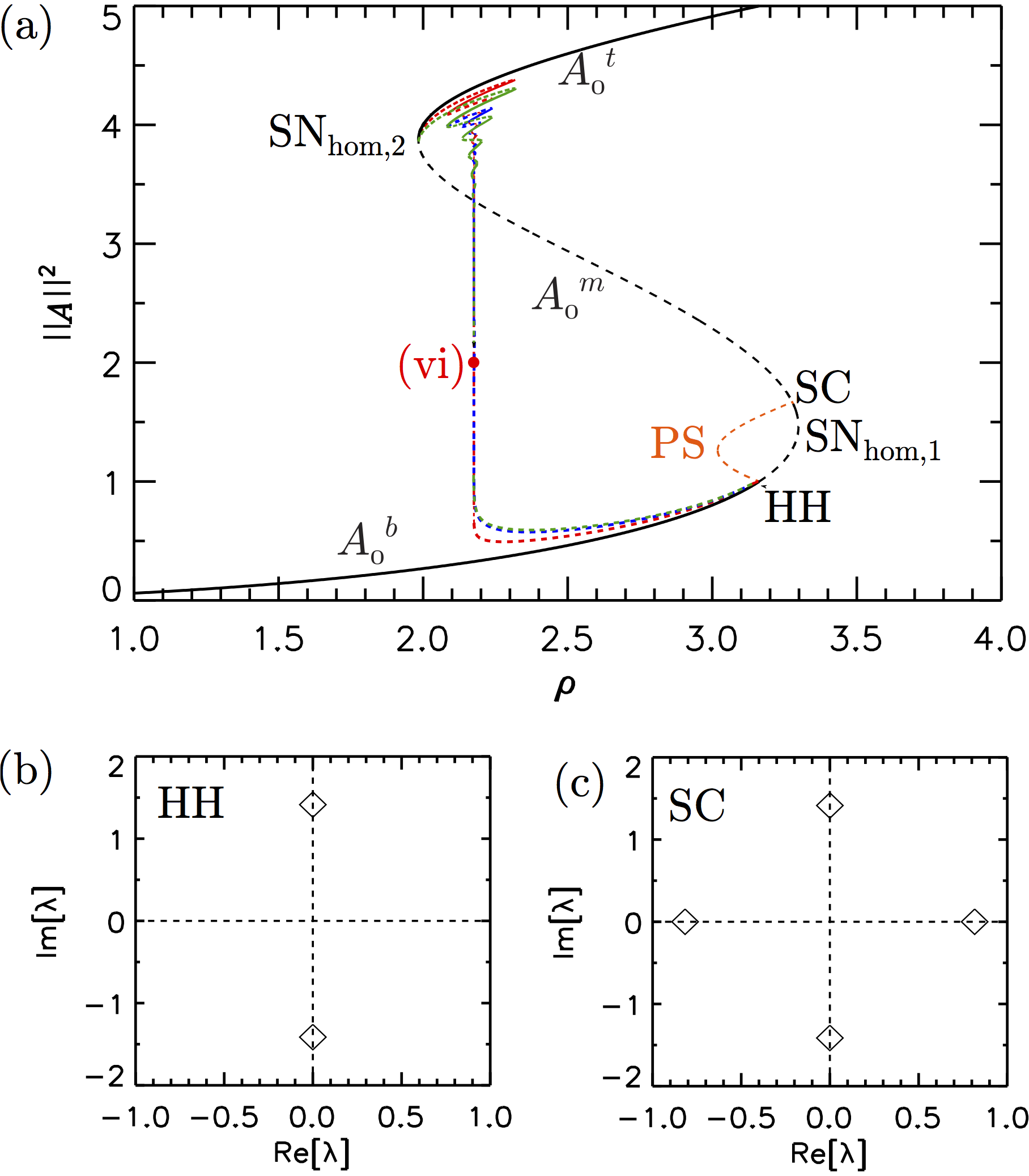}
\captionsetup{justification=raggedright,singlelinecheck=false}
\caption{(Color online) (a) Bifurcation diagram for $\theta=4$ showing collapsed defect-mediated snaking of 1-soliton (red line) and 2-soliton (green line) branches, showing their reconnection with the PS branch (orange line) that bifurcates from HH on $A_0^b$. Temporally stable (unstable) structures are indicated using solid (dashed) lines. Black lines correspond to HSS. Enlargements of panel (a) can be found in Figs.~\ref{newfig7} and \ref{theta4bottom}. (b) The spatial eigenvalues $\lambda$ of $A_0$ at locations HH and SC in (a).}
\label{DarkS_4}
\end{figure}

For $\theta>2$ the saddle node SN$_{hom,1}$ becomes a RTBH point with spatial 
eigenvalues $\lambda_{1,2,3,4}=(0,0,\pm ik_0)$ and homoclinic orbits are 
exceptional \cite{BuYoKn,Kolossovski}. However, in this case this point is 
preceded by a HH bifurcation on $A_0^b$, which gives rise to a branch of PSs. The PSs bifurcate subcritically (Fig.~\ref{DarkS_4})
but remain unstable throughout their existence range, despite the presence of a saddle node.
This is the case for all values of the detuning $\theta$ we explored ($2.3 < \theta < 10$).
Thus no bistability between PS and $A_0^b$ results and no snaking of bright DSs takes
place \cite{Champneys,gomila_Scroggi}. Instead the bright solitons bifurcating from HH 
connect to the dark solitons originating at $\rho=\rho_t$, as we now discuss.

\begin{figure}[t!]
\centering
\includegraphics[scale=1.0]{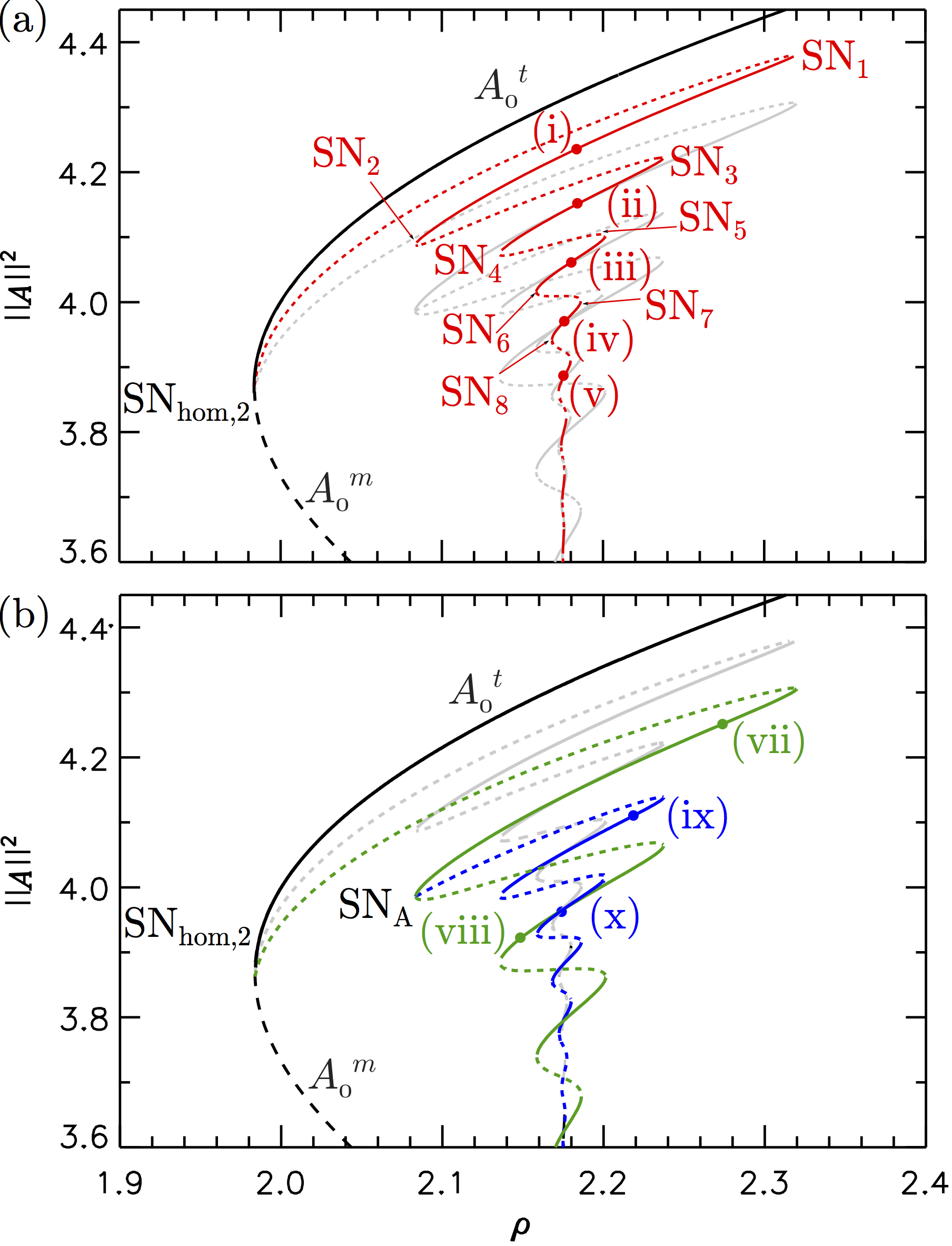}
\captionsetup{justification=raggedright,singlelinecheck=false}
\caption{(Color online) Detail of the 1-soliton (panel (a), red line) and 2-soliton (panel (b), green line) branches in the vicinity of SN$_{hom,2}$ for $\theta=4$. Black lines show the homogeneous states HSS. Panel (b) also shows a family of nonidentical 2-soliton states (blue line) that bifurcate from the saddle node SN$_A$ on the 2-soliton branch and also undergo collapsed defect-mediated snaking. Temporally stable (unstable) structures are indicated using solid (dashed) lines. Profiles corresponding to the labeled locations are shown in Fig.~\ref{LS4}, with details of this process shown in Fig.~\ref{LS4_detail}.}
\label{newfig7}
\end{figure}

\begin{figure}[t!]
\centering
\includegraphics[scale=1]{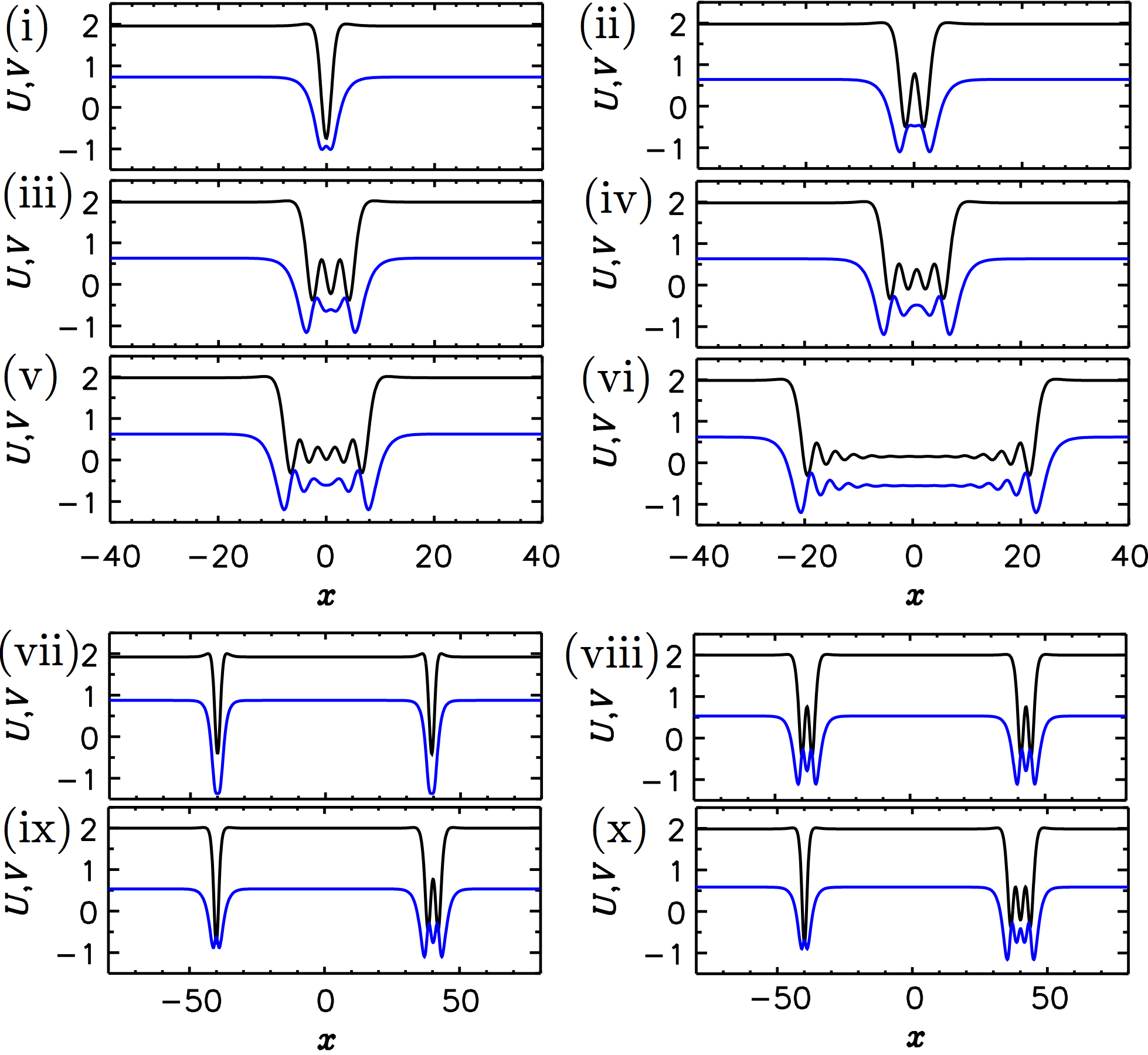}
\captionsetup{justification=raggedright,singlelinecheck=false}
\caption{(Color online) Spatial profiles of the solutions represented in Fig.~\ref{DarkS_4}(a) for $\theta=4$, showing $U(x)$ in black and $V(x)$ in blue. Panels (i)-(vi) correspond to 1-soliton states (red branches in Figs.~\ref{DarkS_4}(a) and \ref{newfig7}(a)), panels (vii)-(viii) to 2-soliton states (green branches in Figs.~\ref{DarkS_4}(a) and \ref{newfig7}(b)) and panels (ix)-(x) to the branch of nonidentical 2-soliton states (blue branch in Fig.~\ref{newfig7}(b)).}
\label{LS4}
\end{figure}
\label{newfig7}

\begin{figure*}[t!]
\centering
\includegraphics[scale=1]{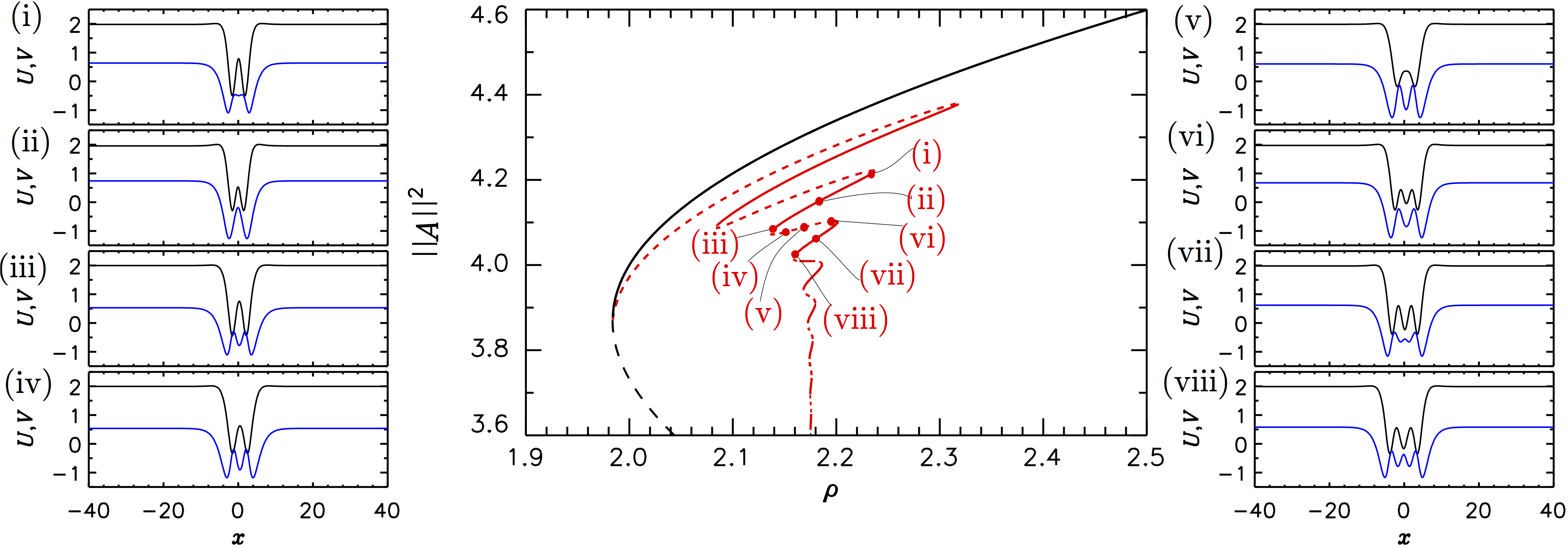}
\captionsetup{justification=raggedright,singlelinecheck=false}
\caption{(Color online) Spatial profiles of the dark solitons near the upper end of the $\theta=4$ 1-soliton branch at locations indicated in the middle panel, showing that the splitting of the central peak (dip) in $(U(x),V(x))$, shown in black and red, respectively, occurs at different locations along the branch.}
\label{LS4_detail}
\end{figure*}

\begin{figure}[t!]
\centering
\includegraphics[scale=1.0]{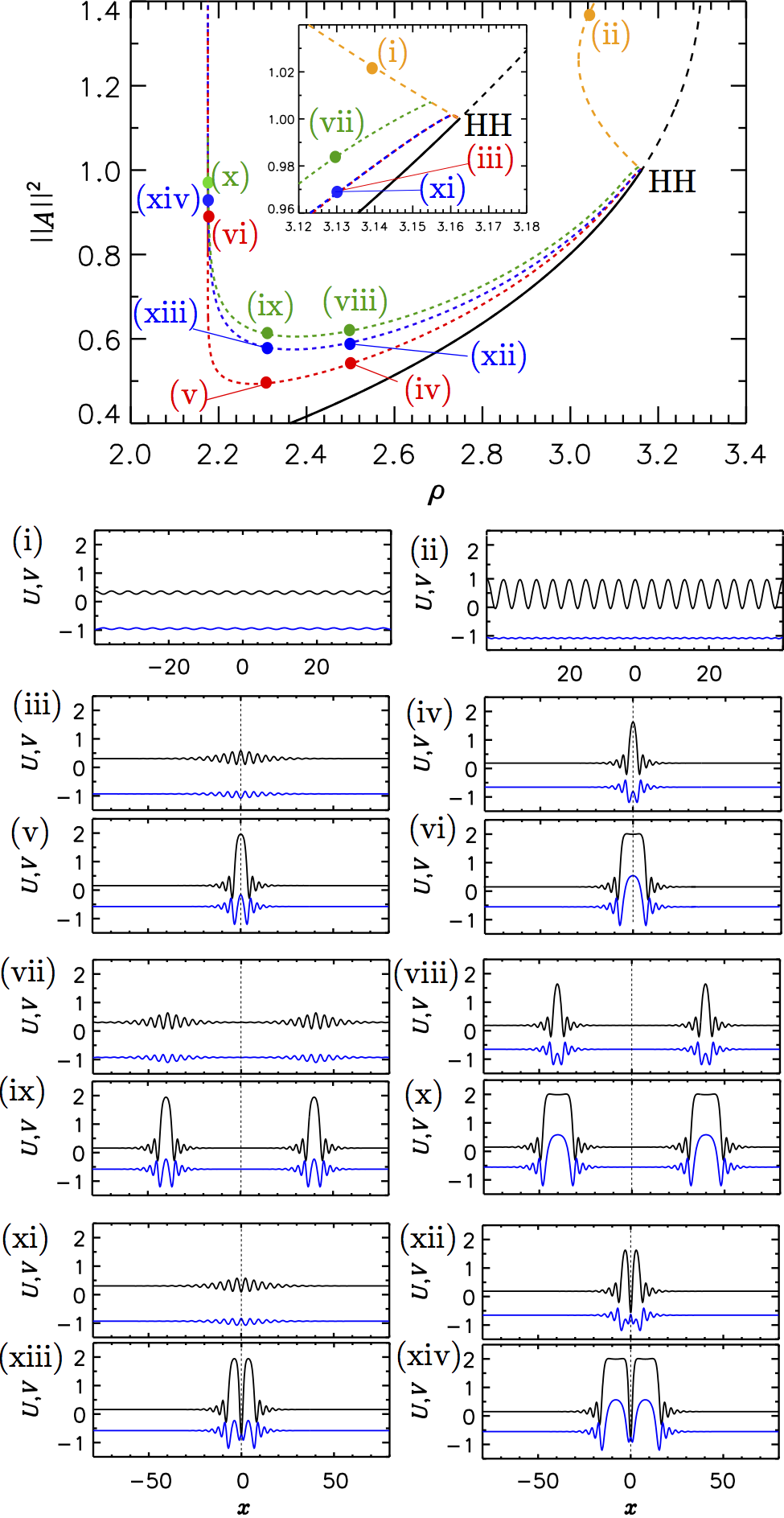}
\captionsetup{justification=raggedright,singlelinecheck=false}
\caption{(Color online) Bifurcation diagram for $\theta=4$ (top panel) showing the bifurcation of the three families
of localized states (bright solitons) from the subcritical PS branch, together with sample solution profiles 
corresponding to the locations indicated in the top panel. States with maxima at $x = 0$ (red line) connect with 
the corresponding branch of dark solitons shown in Figs.~\ref{DarkS_4}(a) and \ref{newfig7}(a) while states with 
minima at $x = 0$ (blue line) connect with the corresponding branch in Fig.~\ref{newfig7}(b). The states shown in 
green consist of two equidistant bright solitons and these connect to the corresponding branch in Fig.~\ref{newfig7}(b).}
\label{theta4bottom}
\end{figure}

\begin{figure*}[t!]
\centering
\includegraphics[scale=1]{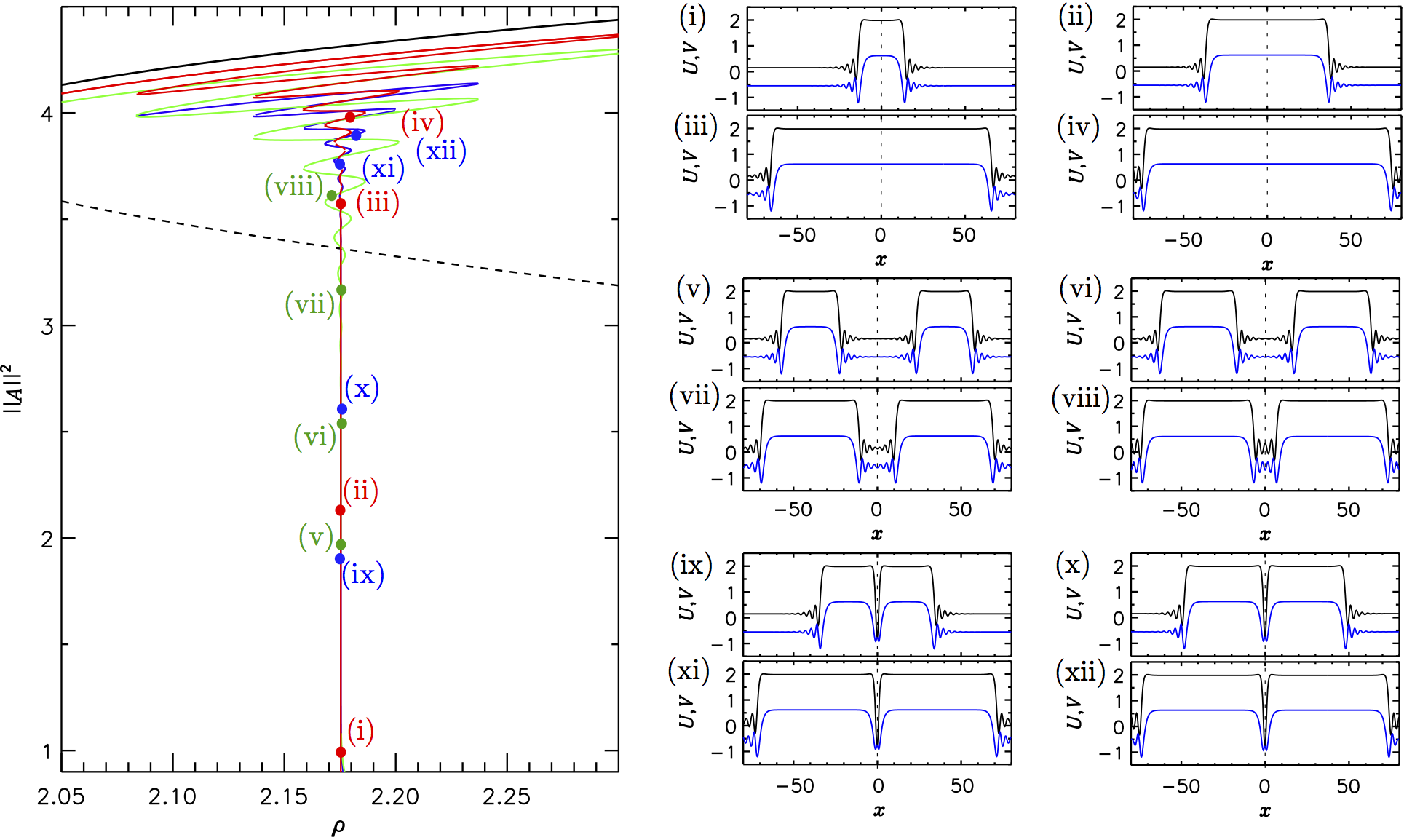}
\captionsetup{justification=raggedright,singlelinecheck=false}
\caption{Details of the profile transformation at $\theta=4$ that changes two nonidentical dark solitons (blue branch in Fig.~\ref{newfig7}(b))
into a bright soliton with a minimum at its center $x=0$, allowing it to connect to the PS state at the same location as the 1-soliton state 
(red branch in Fig.~\ref{newfig7}(a)) which evolves into a bright soliton with a maximum at its center $x=0$. The 2-soliton state consisting of 
two identical equidistant solitons (green branch in Fig.~\ref{newfig7}(b)) also terminates on PS, but at a distinct location.}
\label{extrafigtheta4}
\end{figure*}

Figure~\ref{DarkS_4}(a) shows the bifurcation diagram of the 1-soliton states (red branch) for $\theta=4$ obtained by numerically continuing the analytical
prediction obtained in Eq.~(\ref{sech}) away from SN$_{hom,2}$.  Figure \ref{newfig7}(a) shows a detail of this branch. These states are initially 
unstable but as $\rho$ increases these unstable 1-soliton states grow in amplitude and acquire stability at saddle node SN$_1$. The DS profile
on this segment of the branch is shown in Fig.~\ref{LS4}(i). This solution loses stability at SN$_2$ but starts to develop a spatial oscillation
(SO) in the center; solutions of this type become stable at SN$_3$. An example of the resulting stable solution can be found in Fig.~\ref{LS4}(ii).
This process repeats in such a way that between successive saddle nodes on the left or right a new spatial oscillation is inserted in the center of 
the dark soliton profile and the soliton broadens, decreasing its $L^2$ norm. As a result, as one proceeds down the snaking branch the central 
peak (dip) repeatedly splits. Details of this process are shown in Fig.~\ref{LS4_detail}. The resulting behavior resembles in all aspects the 
phenomenon of defect-mediated snaking described in \cite{BuYoKn_colapsing} except for the exponential shrinking of the region of existence of 
these states as the hole broadens. Consequently we refer to this behavior as collapsed defect-mediated snaking. Numerically the collapse occurs 
at $\rho=\rho_M\approx 2.1753479$. The DSs at this location correspond to broad hole-like states of the type shown in Fig.~\ref{LS4}(v). As in 
Sec.~\ref{Sec::bright} further decrease in the norm signals that the two fronts connecting states $A_0^t$ and $A_0^b$ at $\rho_M$ are starting 
to separate (Fig.~\ref{LS4}(vi)); this process continues, resulting in the bright soliton state shown in Fig.~\ref{theta4bottom}(iv); this state 
is shifted by half the domain width relative to panels (i)-(vi) of Fig.~\ref{LS4}. Thereafter the amplitude of the peak at $x=0$ starts to 
decrease and the 1-soliton branch departs from $\rho_M$, ultimately connecting to the branch of small amplitude PS (Fig.~\ref{theta4bottom}(i)) 
that bifurcates subcritically from HH (see inset in Fig.~\ref{theta4bottom}, top panel).
\begin{figure}[t!]
\centering
\includegraphics[scale=1.0]{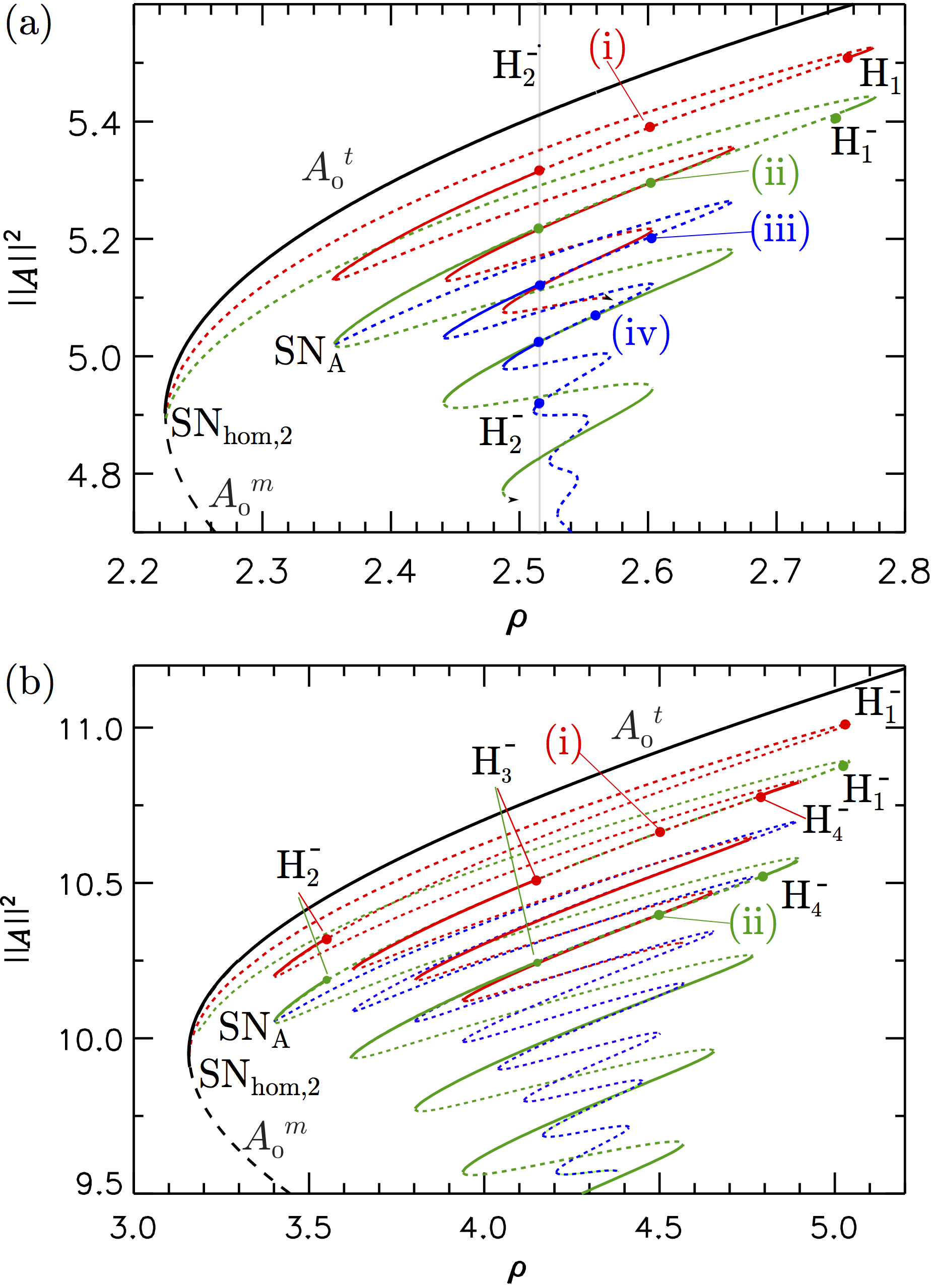}
\captionsetup{justification=raggedright,singlelinecheck=false}
\caption{(Color online) Bifurcation diagram for (a) $\theta = 5$ and (b) $\theta = 10$ showing that the DSs are now unstable within intervals 
between back-to-back Hopf bifurcations. The Hopf bifurcations on the left (H$^-_2$, panel (a)) for the 2-soliton states (green and blue lines) 
coincide with that of the 1-soliton states (red line).}
\label{newfig8}
\end{figure}

Figure~\ref{DarkS_4}(a) also shows the 2-soliton state (green line) that bifurcates from the vicinity of SN$_{hom,2}$ for $\theta=4$ just as in 
the case $\theta=1.95$. For $\theta>2$ this second DS family plays a key role since it is responsible for providing the second of the two 
branches of localized states that are known to be associated with HH. Figures \ref{newfig7}(b), \ref{theta4bottom} and \ref{extrafigtheta4} 
show how this happens. The green branch in Fig.~\ref{newfig7}(b) consists of states with identical equidistant solitons; like the 1-soliton 
states, the 2-soliton states proceed to develop internal oscillations (Figs.~\ref{LS4}(vii)-(viii)). These undergo a symmetry-breaking pitchfork 
bifurcation at SN$_A$ giving rise to a branch of nonidentical solitons (in blue). One of these gradually acquires complex internal structure 
while the other remains unchanged. Figures \ref{LS4}(ix)-(x) show this state at the locations shown in Fig.~\ref{newfig7}(b), while 
Fig.~\ref{extrafigtheta4}(xii) shows a translate of such a 2-soliton state by a quarter of the domain size. Figures \ref{extrafigtheta4}(xii)-(ix)
and \ref{theta4bottom}(xiv)-(xi) show the subsequent evolution of this 2-soliton state into a single wave packet with a minimum at its center 
$x=0$. It is this state that connects to PS at the same location as the corresponding wave packet (red) with a maximum at $x=0$ that originates 
in the 1-soliton state near SN$_{hom,2}$. In contrast, the 2-soliton state that also appears near SN$_{hom,2}$ (green) terminates in a distinct 
bifurcation on PS, as also shown in Fig.~\ref{theta4bottom}. All three branches undergo collapsed defect-mediated snaking inbetween. Evidently 
there are similar branches that bifurcate from other folds on the 2-soliton branch (not shown).

We mention that as the domain length increases the termination point of the 1-soliton (red line) and the nonidentical 2-soliton branch (blue line) migrates towards HH and in the limit of an infinite domain the bright solitons bifurcate from $A_0^b$ simultaneously with the PS, exactly as predicted by the normal form for the spatial Hopf bifurcation with 1:1 resonance \cite{Ioos}. We also mention that, in principle, the Maxwell point $\rho_M$ may collide with the saddle node of the PS branch (see \cite{Ma} for details). However, we have determined that such a collision does not occur in the LLE and that the PS branch remains well-separated from the collapsed snaking branches of dark solitons around $\rho_M$ (at least in the parameter range $2.3 < \theta < 10$). 

We turn, finally, to the structure of the spatial eigenvalues shown in Fig.~\ref{DarkS_4}(b,c). Panel (b) confirms that HH corresponds to a Hamiltonian-Hopf bifurcation in space. Panel (c) shows that at the termination point of the PS branch the HSS state $A_0^m$ has 2 purely real and 2 purely imaginary spatial eigenvalues, indicating that SC corresponds to a global bifurcation in space and not a local bifurcation. Both HH and SC are formed in the process of unfolding the spatially reversible QZ bifurcation that takes place at SN$_{hom,1}$ when $\theta=2$.

\subsection{Soliton location in the $(\theta,\rho)$ plane}

Tracking each bifurcation point in the bifurcation diagram as a function of $\theta$ we 
obtain the $(\theta,\rho)$ parameter plane shown in Fig.~\ref{parameter_space}. The 
green solid line represents a BD transition for $\theta<2$ that turns into a HH 
bifurcation for $\theta>2$. The saddle-node bifurcations determine the regions 
of existence of the different dark solitons shown previously. With increasing 
$\theta$ the region of existence of these states becomes broader (Fig.~\ref{newfig8}(a,b)).
In contrast, when $\theta$ decreases the branches of solutions with several SO progressively
shrink, disappearing in a series of cusp bifurcations C$_{1}$,...,C$_4$, as shown in 
Fig.~\ref{parameter_space}. 

We distinguish four main dynamical regions, labeled I to IV in the phase diagram 
in Fig.~\ref{parameter_space}, on the basis of the existence of HSS and dark DSs:

\begin{itemize}
\item Region I: The bottom HSS $A_0^b$ is stable. No dark DSs or top HSS $A_0^t$ exist. This region spans the parameter space $\rho < \rho_{BD}$ for $\theta < \sqrt{3}$ and $\rho < \rho_{SN_{hom,2}}$ for $\theta > \sqrt{3}$.

\item Region II: The bottom HSS $A_0^b$ and top HSS $A_0^t$ coexist and are both stable. No dark DSs are found. This region spans the parameter space $ \rho_{SN_1} < \rho < \rho_{SN_{hom,1}}$ for $\theta > \sqrt{3}$.

\item Region III: The top HSS $A_0^t$ is stable. No dark DSs or bottom HSS $A_0^b$ exist. This region spans the parameter space $\rho > \rho_{BD}$ for $\theta < \sqrt{3}$ and $\rho > \rho_{SN_{hom,1}}$ for $\theta > \sqrt{3}$.

\item Region IV: The bottom HSS $A_0^b$ and top HSS $A_0^t$ are stable and coexist with (possibly unstable) dark DSs. This region spans the parameter space $\rho_{SN_{hom,2}} < \rho < \rho_{SN_{1}}$ for $\theta > \sqrt{3}$.

\end{itemize}

Region IV is the main region of interest in this work. It can be further subdivided to reflect the locations of different types of DSs. In 
the next Section, we refer to the region between SN$_1$ and SN$_2$, i.e., the region of existence of 1-SO dark solitons, as subregion IV$_{1}$. 
Similarly, subregion IV$_{2}$ corresponds to 2-SO dark solitons between SN$_3$ and SN$_4$ and so on. While both HSS are stable in region IV, 
the stability of dark DSs in the various subregions depends on the parameter values ($\theta$,$\rho$) as discussed next.

\begin{figure}[t!]
\centering
\includegraphics[scale=0.9]{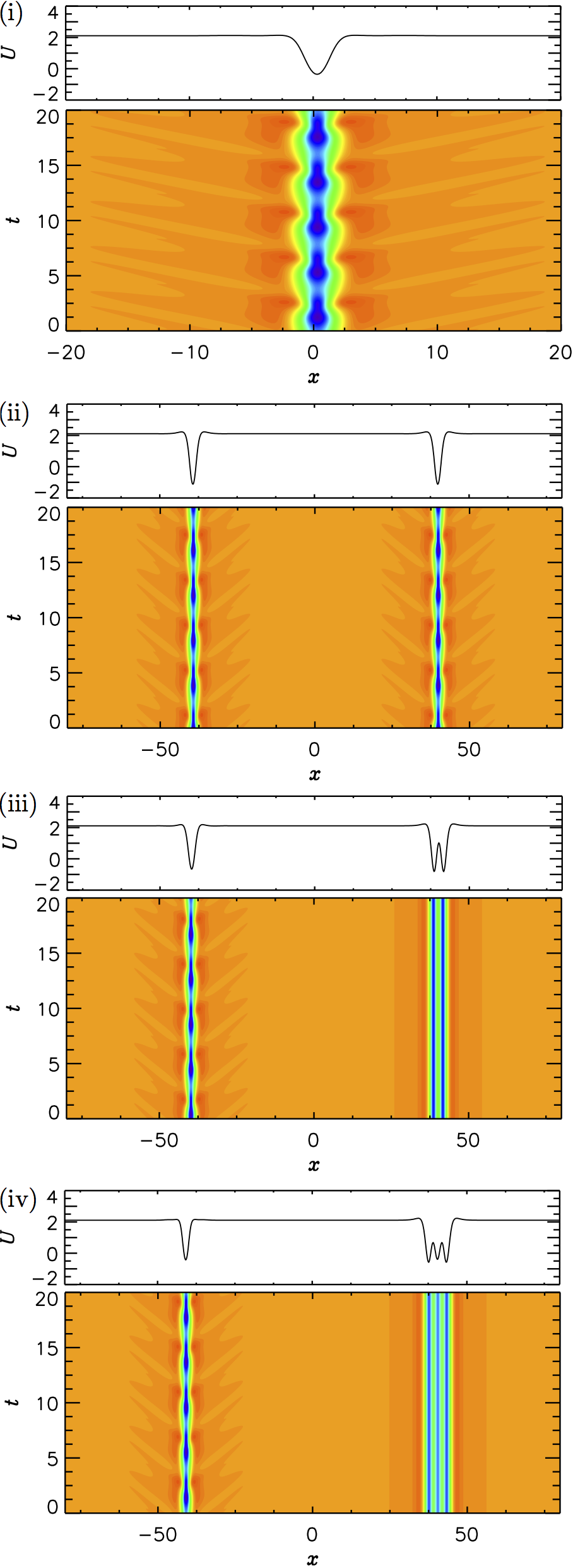}
\captionsetup{justification=raggedright,singlelinecheck=false}
\caption{(Color online) (i) Oscillatory 1-soliton state, (ii) oscillatory 2-soliton state, (iii) a bound state of 
an oscillating and a stationary dark soliton, all computed for $\theta=5$, $\rho=2.6$. (iv) A similar
state to panel (iii) but for $\theta=5$, $\rho=2.56$. The solutions are represented in a space-time plot of
$U(x,t)$  with time increasing upwards. The profile at the final instant, $t=20$, is shown above each space-time plot.}
\label{Osci_DarkS_5}
\end{figure}
\begin{figure}[t!]
\centering
\includegraphics[scale=0.9]{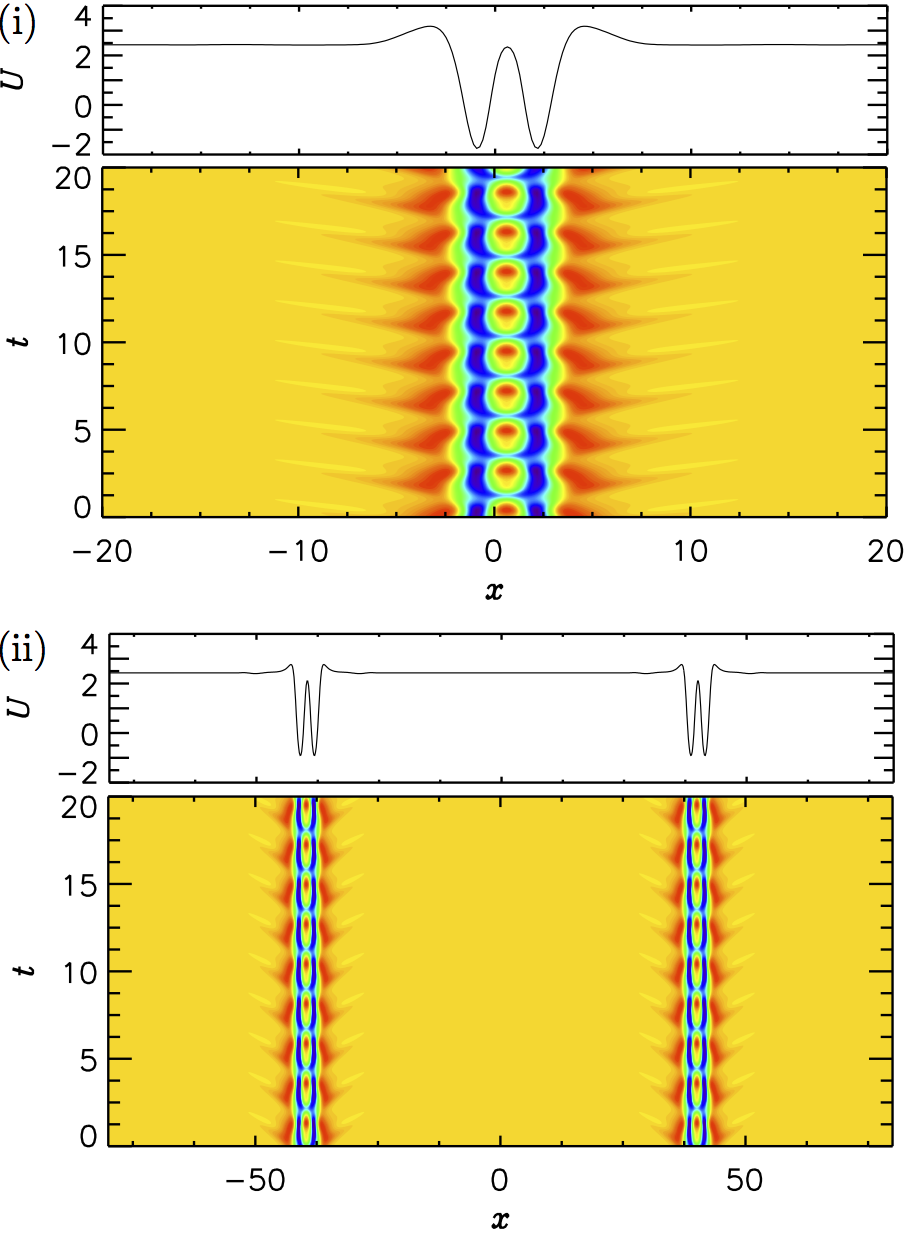}
\captionsetup{justification=raggedright,singlelinecheck=false}
\caption{(Color online) (i) Oscillatory 1-soliton state, and (ii) oscillatory 2-soliton state, when $\theta=10$, $\rho=4.5$. The solutions are represented in a space-time plot of $U(x,t)$ with time increasing upwards. The profile at the final instant, $t=20$, is shown above each space-time plot.}
\label{Osci_DarkS_10}
\end{figure}

\section{Oscillatory and chaotic dynamics}\label{Sec::osci}

We have seen that the range of values of the parameter $\rho$ within which one finds dark solitons increases rapidly with increasing detuning $\theta$ although the interval with stable stationary dark solitons is reduced by the presence of oscillatory instabilities that set in as $\theta$ increases (Fig.~\ref{newfig8}). These intervals of instability open up on the stable portions of the collapsed snaking branches, between pairs of supercritical Hopf bifurcations on either side. Consequently these instabilities lead to stable temporal oscillations resembling breathing of the individual solitons. To characterize the resulting dynamics we combine here linear stability analysis in time with direct integration of the LLE. We also compute secondary bifurcations of time-periodic states and point out that in appropriate regimes the LLE exhibits dynamics that are very similar to those exhibited by excitable systems. 

\begin{figure}[t!]
\centering
\includegraphics[scale=1]{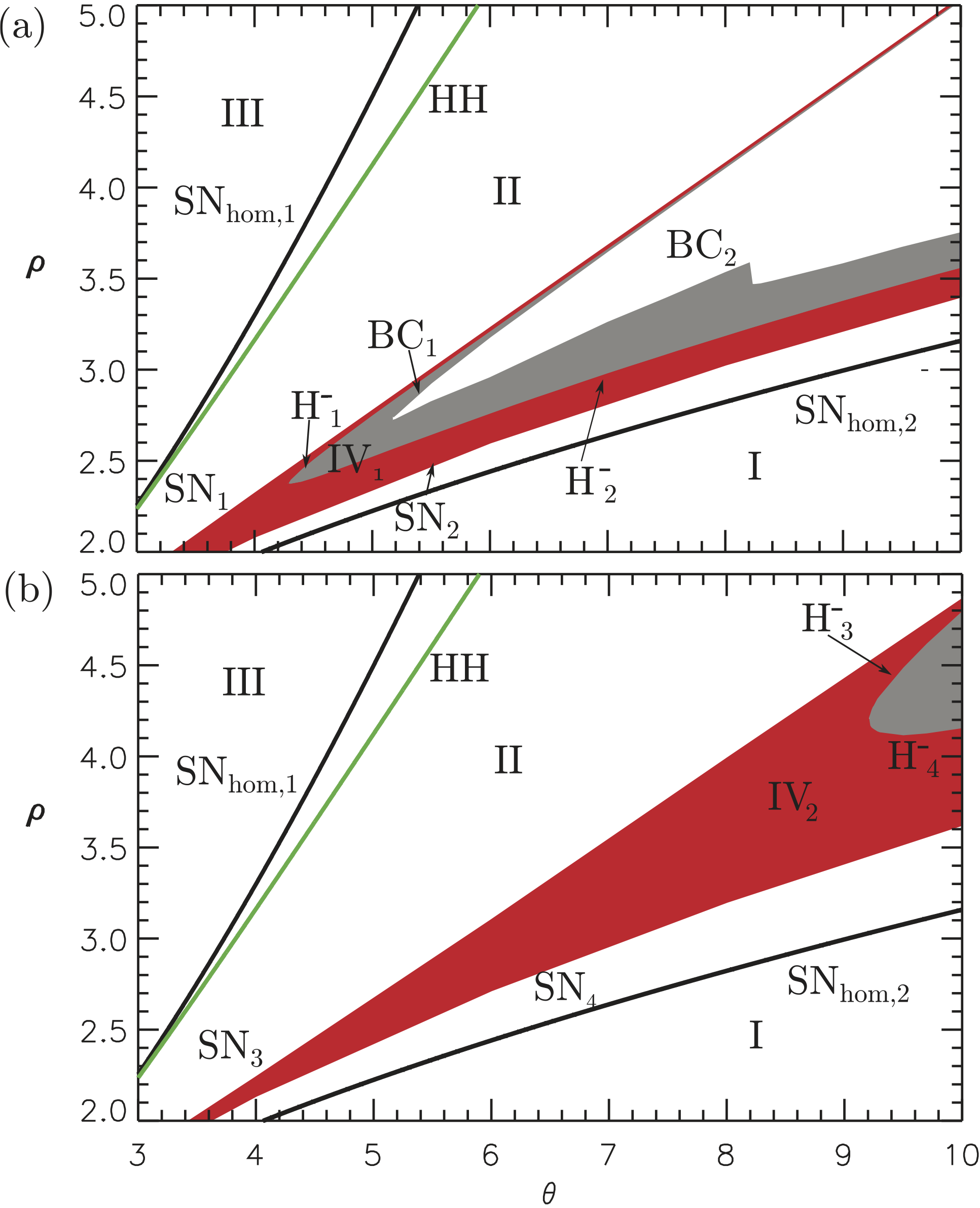}
\captionsetup{justification=raggedright,singlelinecheck=false}
\caption{(Color online) The $(\theta,\rho)$ parameter space for normal dispersion 
($\nu=-1$) showing the region of existence of (a) the 1-SO dark soliton and (b) 
the 2-SO dark soliton. The different bifurcations are labeled, with $H^-_j$ indicating
a supercritical Hopf bifurcation at location $H_j$. The red (gray) 
region corresponds to stable stationary (oscillatory) dark DSs.}
\label{parameter_space_H}
\end{figure}

Figures~\ref{newfig8}(a) and (b) show that for both $\theta=5$ and $\theta=10$ 
the single dark soliton becomes unstable in a supercritical Hopf bifurcation (H$^-_1$) 
leading to an oscillatory state. Figure~\ref{Osci_DarkS_5}(i) shows the resulting
state at location (i) in Fig.~\ref{newfig8}(a). The temporal oscillations disappear
upon further decrease in $\rho$ and do so in a reverse supercritical Hopf at H$^-_2$, 
thereby restoring the stability of the single dark soliton. For larger values of
$\theta$ this behavior not only persists but the soliton with 2 spatial oscillations 
(SO) also exhibits temporal oscillations between two back-to-back Hopf bifurcations 
(Fig.~\ref{newfig8}(b)). An example of such oscillatory 2-SO dark soliton is shown 
in Fig.~\ref{Osci_DarkS_10}(i). 

Figure ~\ref{Osci_DarkS_5}(ii) shows the corresponding oscillation of the 2-soliton state
for $\theta=5$ at location (ii) in Fig.~\ref{newfig8}(a). The solitons oscillate in phase 
but in a nonsinusoidal manner. Figures~\ref{Osci_DarkS_5}(iii)-(iv) show oscillations of 
a bound state of two nonidentical dark solitons at locations (iii) and (iv) in 
Fig.~\ref{newfig8}(a). In these states the simple dark soliton on the left oscillates in 
a periodic fashion while the structured dark soliton on the right remains essentially 
time-independent. Figure~\ref{Osci_DarkS_10}(ii) shows a periodic oscillation of a 
2-soliton state for $\theta=10$ corresponding to location (ii) in Fig.~\ref{newfig8}(b). 
The individual solitons are structured and oscillate as in panel (i). Once again, both 
oscillate in phase. 

We can complete the parameter space shown in Fig.~\ref{parameter_space} by adding the 
curves corresponding to the oscillatory instabilities at H$^-_1$ and H$^-_2$. 
Figure~\ref{parameter_space_H} shows the parameter space with the curves corresponding 
to the temporal instabilities of the 1-SO and 2-SO dark solitons included; the saddle nodes 
of the remaining dark solitons are omitted in order to give a clearer understanding of 
this behavior.
\begin{figure}[t!]
\centering
\includegraphics[scale=1]{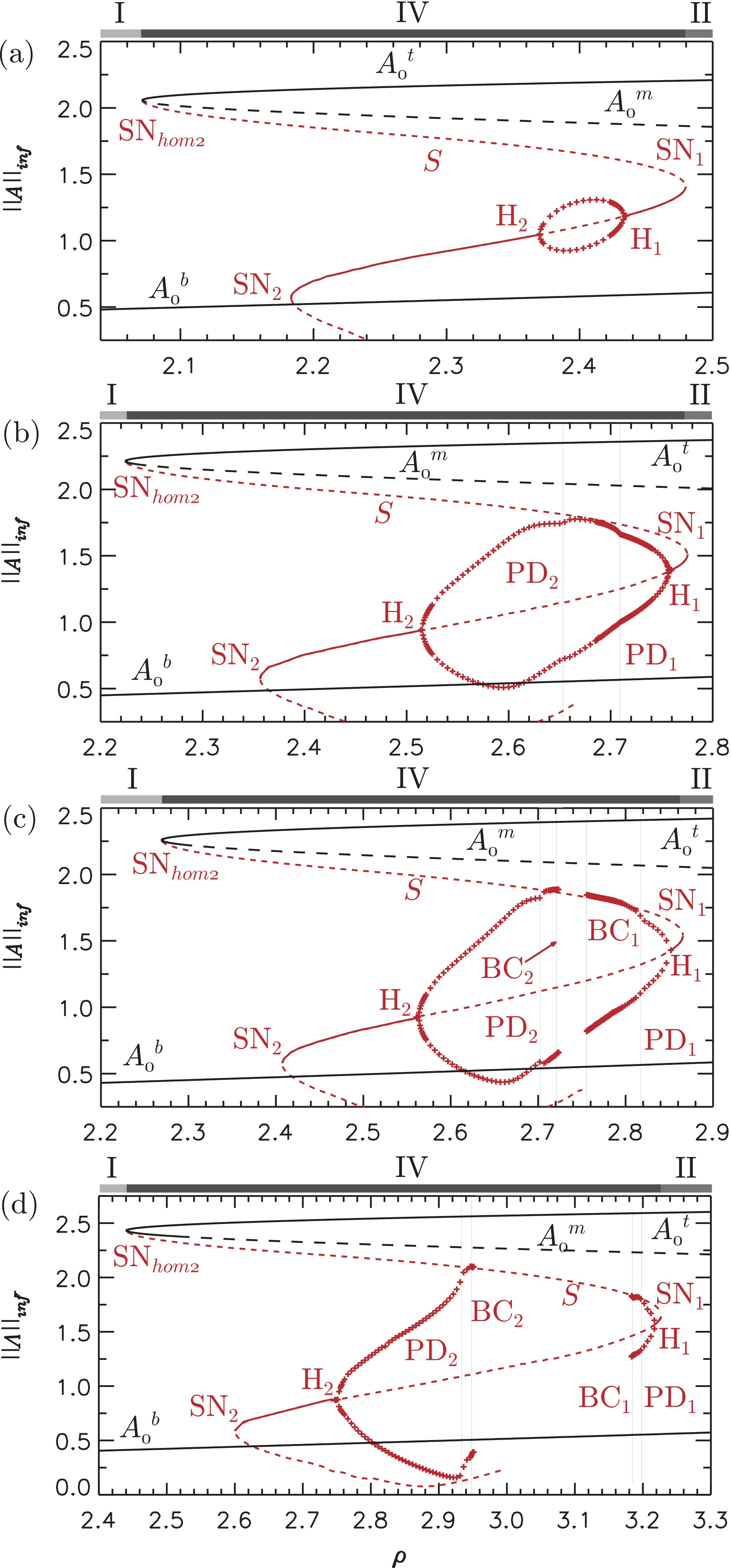}
\captionsetup{justification=raggedright,singlelinecheck=false}
\caption{Bifurcation diagrams corresponding to different slices of the parameter space in Fig.~\ref{parameter_space_H} plotted using 
$||A||_{\rm inf}$ as a measure of the amplitude. Solid (dashed) lines correspond to stable (unstable) 
structures, and red (black) colors correspond to 1-SO dark DSs (HSS). The red crosses represent maxima and
minima of the amplitude of the oscillatory dark DSs. The gray labeled bars above each panel show the extent 
of the regions I, II, and IV. (a) $\theta=4.6$, (b) $\theta=5$, (c) $\theta=5.2$, (d) $\theta=5.5$.}
\label{Hopfs}
\end{figure}
\begin{figure}[t!]
\centering
\includegraphics[scale=1]{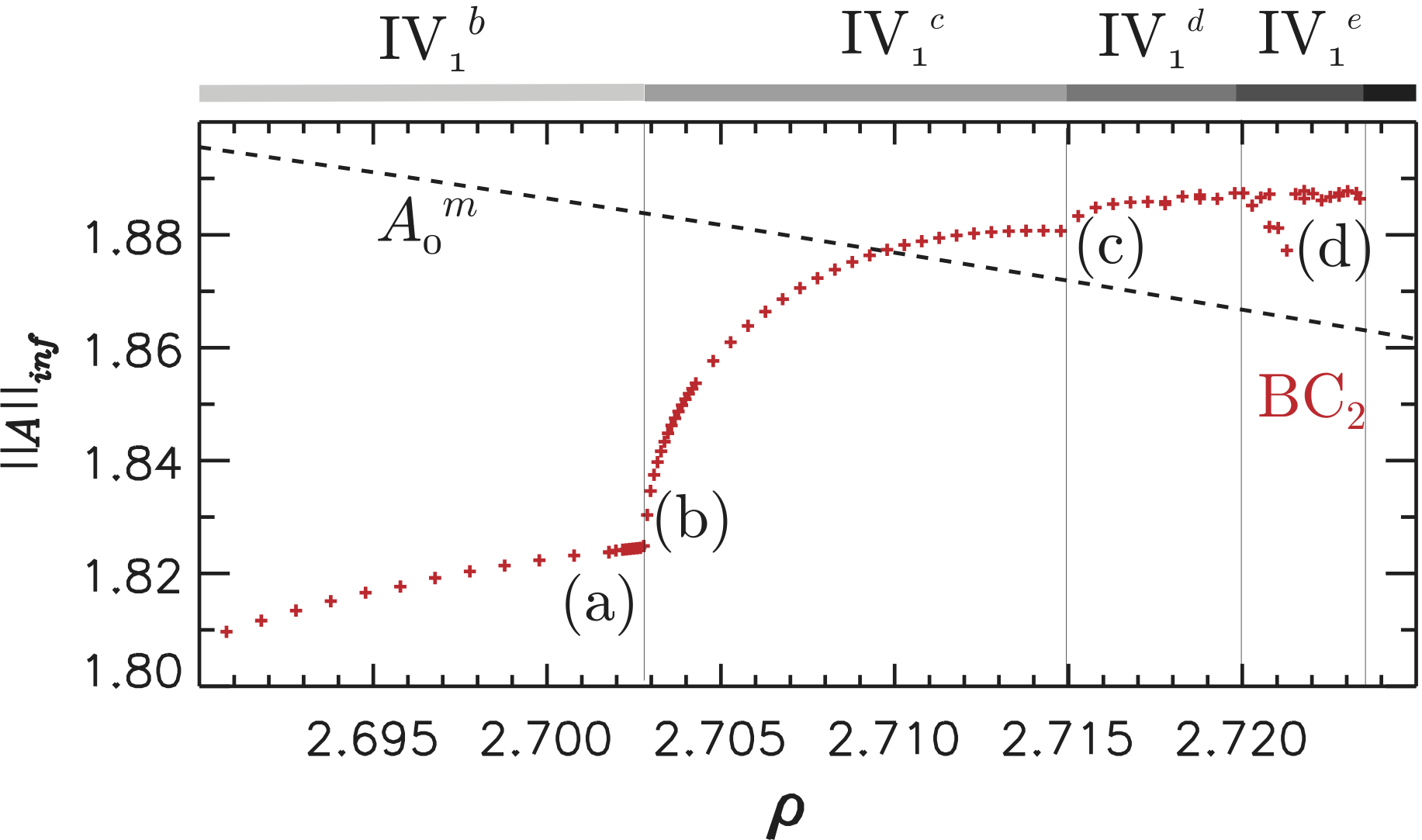}
\captionsetup{justification=raggedright,singlelinecheck=false}
\caption{Detail of the bifurcation diagram of Fig.~\ref{Hopfs}(c) for $\theta=5.2$ close to the BC$_1$ point. Vertical lines separate period 1 oscillations (region IV$_1^b$), period 2 oscillations (region IV$_1^c$), period 4 oscillations (region IV$_1^d$) and temporal chaos in region IV$_1^e$. Lines and markers in red (black) correspond to dark DSs (HSS). Labels from (a) to (d) correspond to the dynamics shown in Fig.~\ref{atract}.} 
\label{detalleH}
\end{figure}
Bifurcation lines separating different dynamical regimes are labeled according 
to Fig.~\ref{newfig8}. With increasing $\theta$ the Hopf 
bifurcation H$^-_1$ of the single dark DS approaches SN$_1$ and we see 
that both lines are almost tangent although, for the parameter values presented, 
they do not meet. The same scenario repeats for the Hopf bifurcation H$^-_3$ 
of the 2-SO state.

\begin{figure*}[t!]
\centering
\includegraphics[scale=1]{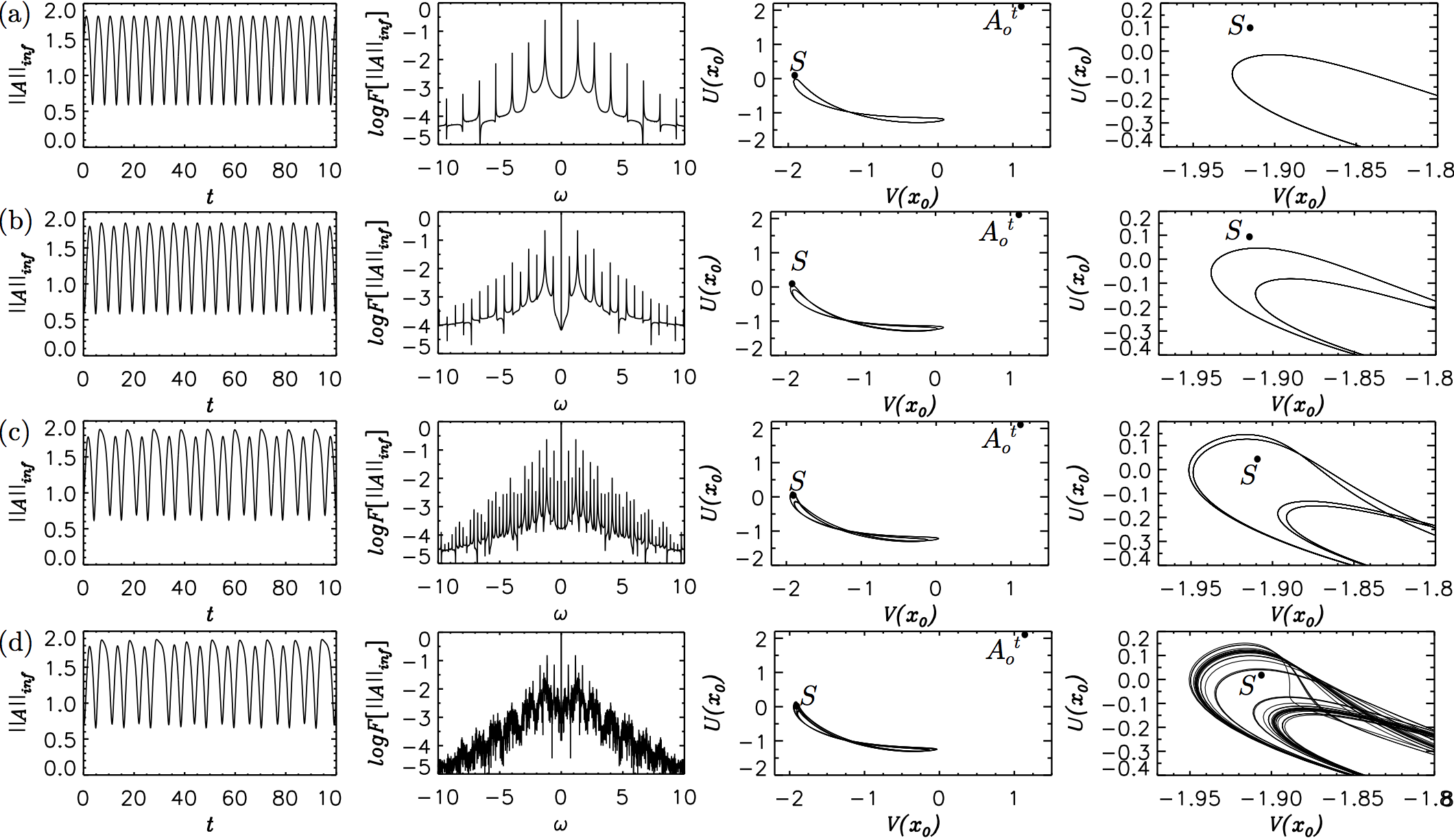}
\captionsetup{justification=raggedright,singlelinecheck=false}
\caption{Route to temporal chaos for $\theta=5.2$. Panels (a)--(d) represent the transition from (a) period 1 oscillations to (d) temporal chaos, corresponding to the labels in Fig.~{\ref{detalleH}}. From left to right: temporal trace of $||A||_{\rm inf}$, its frequency spectrum that allows us to differentiate between the different types of temporal periodicity, a portion of the phase-space containing $A_0^t$, $S$ and the periodic attractors, and a zoom of the latter where we can appreciate the proximity of $S$ to the cycle. (a) $ \rho=2.70248$ (period 1), (b) $\rho=2.70358$ (period 2), (c) $\rho=2.71528$ (period 4), (d) $\rho=2.72178$ (temporal chaos).} 
\label{atract}
\end{figure*}

This scenario can be better understood by looking at Fig.~\ref{Hopfs} where several 
slices of Fig.~\ref{parameter_space_H} at different values of $\theta$ are 
shown. We choose to plot $||A||_{\inf}:={\rm min}(|A|)$ instead of the $L^2$ norm to improve 
the clarity of the bifurcation diagram, and denote the maximum and minimum amplitude of the 
oscillatory DSs using crosses. The diagram in Fig.~\ref{Hopfs}(a) corresponds 
to a cut of Fig.~\ref{parameter_space_H} at $\theta=4.6$. At this $\theta$ value the
oscillatory state bifurcates from H$^-_1$, grows in amplitude as $\rho$ 
decreases, before reconnecting to the stationary DS at H$^-_2$ in a reverse Hopf 
bifurcation. For larger $\theta$, the amplitude of the limit cycle between 
H$^-_1$ and H$^-_2$ increases, and at some point the cycle undergoes a
period-doubling (PD) bifurcation, starting a route to a chaotic attractor. This 
happens already at $\theta=5$ as can be seen in Fig.~\ref{Hopfs}(b). At 
$\theta=5.2$ (Fig.~\ref{Hopfs}(c)) the chaotic attractor touches the saddle 
branch $S$ corresponding to unstable dark solitons and disappears through a 
boundary crisis (BC) \cite{Hilborn}. Let us discuss this process in detail for 
the cycle emerging from H$^-_2$ (the case of H$^-_1$ is analogous). In 
Fig.~\ref{detalleH} we show a zoom of the diagram in Fig.~\ref{Hopfs}(c) close 
to BC$_2$ and in Fig.~\ref{atract} a series of panels characterizing the cycle 
at different values of $\rho$ is shown. From left to right we show a series of 
time traces corresponding to the temporal evolution of the minima of the 
soliton, i.e., $||A||_{\inf}$, the Fourier transform of these time traces, a 
two-dimensional phase space projection onto $(U(x_0,t),V(x_0,t))$, $x_0$ being 
the position of the center of the structure, and a zoom of the phase space. 
Panel (a) in Fig.~\ref{atract} corresponds to the situation at $ \rho=2.70248$ in 
Fig.~\ref{detalleH} labeled with (a). As we can see in the time trace and in the 
frequency spectrum, the cycle has a single period. In the phase space shown in 
Fig.~\ref{atract} we observe a fixed point corresponding to $A_0^t$, a saddle 
point corresponding to the unstable dark soliton denoted by $S$ and a periodic 
orbit corresponding to the cycle. For this value of $\rho$ the saddle $S$ is far 
from the cycle. For $ \rho=2.70358$ (panel (b) corresponding to label (b) in 
Fig.~\ref{detalleH} the time trace and the spectrum reveal that the cycle has 
period two as can also be discerned from the phase space projection. In 
Fig.~\ref{atract}(c) for $\rho=2.71528$ the cycle has just suffered another 
period-doubling resulting a cycle with period four. Finally, 
Fig.~\ref{atract}(d) shows the situation for $\rho=2.72178$, where the cycle has 
become a chaotic attractor. At this parameter value the system is very close to the 
boundary crisis BC$_2$ as can appreciated from the near tangency between $S$ 
and the chaotic attractor. Once $S$ touches the attractor, the latter disappears 
and only $A^t_0$ and $A_0^b$ remain as attractors of the system. The same occurs
to the cycles appearing at H$^-_1$. Using time simulations we were able to 
estimate the position of the boundary crises BC$_1$ and BC$_2$ in 
parameter space, labeled in Fig.~\ref{parameter_space_H}(a).
From Fig.~\ref{Hopfs}(c) to Fig.~\ref{Hopfs}(d) we 
can see that at the same time as BC$_1$ moves toward H$^-_1$, H$^-_1$ itself 
approaches SN$_{1}$ and therefore that the region of existence of oscillatory
DSs shrinks. This behavior can also be seen in Fig.~\ref{parameter_space_H}(a). 

At this point we can differentiate five main dynamical subregions related to region IV$_1$, i.e., the 1-SO dark soliton, namely:
\begin{itemize}
 \item IV$_1^a$: The 1-SO dark soliton is stable
 \item IV$_1^b$: The soliton oscillates with a single period
 \item IV$_1^c$: The soliton oscillates with period two
 \item IV$_1^d$: The soliton oscillates with period four
 \item IV$_1^e$: Region of temporal chaos bounded by a boundary crisis (BC$_2$). 
\end{itemize}
The region IV$_2$ of 2-SO dark solitons has the same sequence of subregions IV$_2^a$,...,IV$_2^e$, etc. 
 
\begin{figure*}[t!]
\centering
\includegraphics[scale=1]{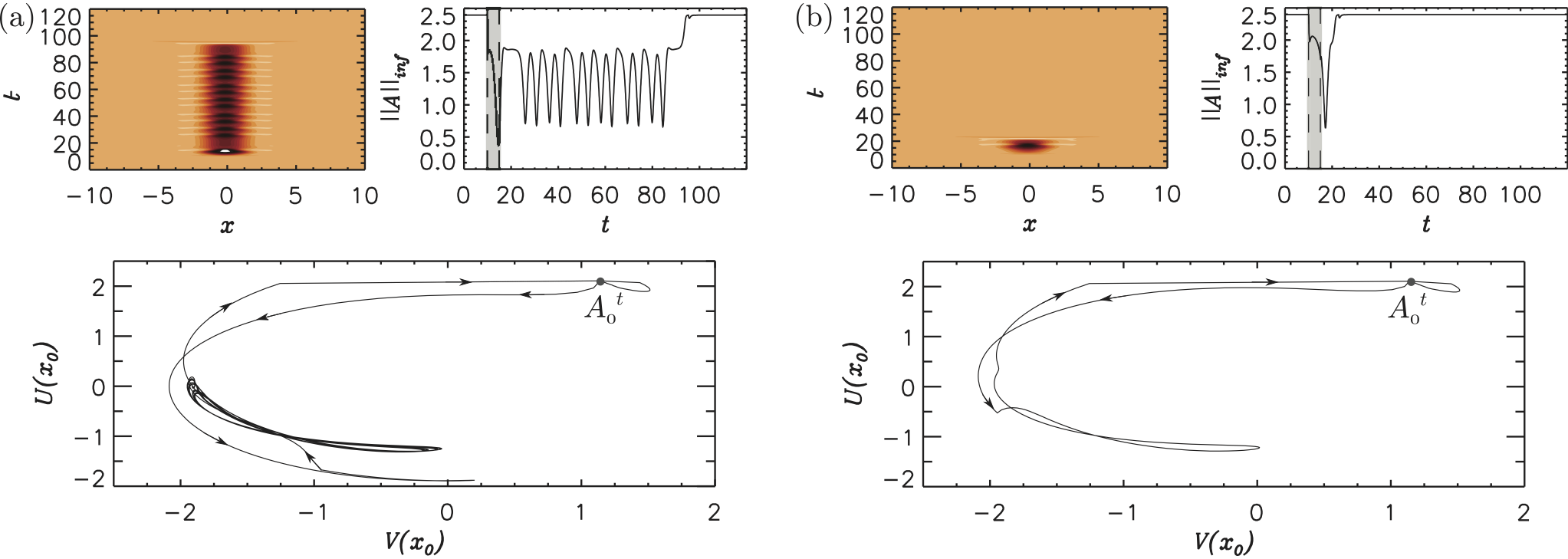}
\captionsetup{justification=raggedright,singlelinecheck=false}
\caption{(Color online) Chaotic transient dynamics for $\theta=5.2$: (a) A chaotic transient is generated when $A_0^t$ is temporally perturbed with a Gaussian perturbation of height $h=-2.55$ (see gray area in time traces); (b) a similar excursion for $h=-3.4431$. In both (a) and (b) the top left panels represent space-time plots of the temporal evolution of the field $U(x,t)$, the top right panels show the time series of the norm $||A||_{\inf}$ and the bottom panels a projection of the phase space trajectory.} 
\label{excita}
\end{figure*}

Close to BC$_2$ (respectively, BC$_1$) the system can exhibit behavior reminiscent of excitability \cite{gomila}. Here the stable manifold of the saddle soliton $S$ acts as a separatrix or threshold in the sense that perturbations of $A_0^t$ across that threshold do not relax immediately to $A^t_0$ but lead first to a large excursion in phase space before relaxing to $A_0^t$. In this case the excursion corresponds to what is known as a chaotic transient, where the system exhibits transient behavior reminiscent of the chaotic attractor at lower values of $\rho$ \cite{leo-lendert, Ott}. In Figs.~\ref{excita}(a) and (b) we show two examples of this kind of transient dynamics. We choose a value of $\rho$ close to BC$_2$, namely $\rho=2.7235$, and modify the parameter $\rho$ for a brief instant using a Gaussian profile of width $\sigma$ and height $h$ using the instantaneous transformation $\rho\mapsto \rho+h(t)\exp[-(x-L/2)^2/\sigma^2]$, where $\rho=2.7235$ and $\sigma=0.781250$ with $h(t)=-2.55$ for $10\le t\le 15$ and $h=0$ elsewhere \cite{Parra_Rivas_drift_defect}. As shown in Fig.~\ref{excita}(a) such a perturbation of $A_0^t$ allows the system to explore the chaotic attractor before returning to the rest state. In contrast, in Fig.~\ref{excita}(b) the system explores just one loop of the cycle before returning to the rest state. 

\section{Concluding remarks}\label{Sec::conclusions}

In this work we have presented a comprehensive overview of the dynamics of the LLE in the normal dispersion regime. The bifurcation structure of dark dissipative solitons (DSs), their stability and the regions of their existence were determined. Three families of dark solitons, the 1-soliton and two different types of 2-soliton states, located on three intertwined branches undergoing collapsed snaking in the vicinity of the same Maxwell point, were identified. The 1-soliton states bifurcate from the top left fold of an S-shaped branch of spatially homogeneous states and terminate either on the lower homogeneous steady state (HSS) branch in a Hamiltonian-Hopf (HH) (equivalently, modulational instability) or at the bottom right fold, depending on the detuning parameter $\theta$. On a periodic domain of finite spatial period, these bifurcations are slightly displaced from the folds, and in the case of the HH bifurcation to finite amplitude on the branch of periodic states created in this bifurcation. The 2-soliton states consisting of a pair of identical equidistant solitons in the domain follow a similar branch but branch off the HSS farther from the folds. This is a finite size effect: these states behave like the 1-soliton states on a periodic domain with half the domain length. The third branch consists of a pair of nonidentical solitons and plays a key role: this branch bifurcates from the branch of identical 2-soliton states in a pitchfork bifurcation; as one follows this branch to lower $L^2$ norm these states undergo a remarkable metamorphosis into a bright soliton with a minimum at its center that allows it to terminate on the periodic states created in the HH bifurcation at the same location as the 1-soliton states, as demanded by theory. The details of this transition are captured in Figs.~\ref{theta4bottom} and \ref{extrafigtheta4}.

At yet higher values of the detuning parameter $\theta$ we found that the localized states undergo oscillatory instabilities, and at a certain point a period-doubling bifurcation initiates a period-doubling cascade into chaos. We have used this observation to determine the regions in parameter space where different stationary and dynamical states coexist. 

We have shown that the bifurcations that organize the spatial dynamics undergo an important transition at a Quadruple-Zero (QZ) point, which occurs at $(\theta,\rho) = (2,\sqrt{2})$. Here, in the normal dispersion regime, the Belyakov-Devaney (BD) transition turns into an HH bifurcation as the detuning $\theta$ increases through $\theta=2$. For $\theta > 2$ a spatially periodic pattern bifurcates subcritically from the bottom homogeneous state at this HH bifurcation. These periodic solutions were found to be unstable, and hence no stable bright DSs were found. However, the saddle-node bifurcation of the top homogeneous solution remains a reversible Takens-Bogdanov (RTB) bifurcation for all $\theta > \sqrt{3}$. This observation explains the existence of multiple families of dark DSs in this regime, and their organization in the so-called \textit{collapsed snaking} structure \cite{BuYoKn,BuYoKn_colapsing}. As mentioned, these dark DSs undergo various dynamical instabilities for larger values of the detuning $\theta$.

The bifurcation scenario is largely reversed in the case of anomalous dispersion, where the same QZ point plays an equally important role, but now the HH bifurcation turns into a BD bifurcation when $\theta > 2$ \cite{Parra_Rivas_2,Goday_chembo}. Moreover, the top homogeneous solution is now always unstable and the upper fold never corresponds to a RTB bifurcation. This reverse character of the bifurcation points has important consequences. First, dark DSs no longer exist, although the inclusion of additional, higher order dispersion can stabilize the top homogeneous solution and hence lead to stable DSs \cite{Gelens_darkDS}. Second, for $41/30<\theta<2$, a stable periodic solution coexists with the stable bottom homogeneous solution giving rise to bright DSs that are organized in a homoclinic snaking structure \cite{gomila_Scroggi,Champneys}. For $\theta>2$, however, the snaking structure of such bright DSs breaks down, as will be reported elsewhere. Finally, despite these differences in the regions of existence of dark and bright DSs in the normal vs. anomalous dispersion regime, the temporal dynamics of the existing solutions are very similar at higher values of the detuning $\theta$. Here, for normal dispersion, we reported the existence of oscillatory and chaotic dynamics of dark DSs as the detuning is increased. The same dynamical instabilities have been observed in the case of anomalous dispersion at high values of $\theta$, but this time for bright DSs \cite{leo-lendert,Parra_Rivas_2}. This suggests that the unfolding of the dynamics can be related to the same type of bifurcation point in both cases. 

Owing to the strong correspondence between Kerr temporal solitons and frequency combs (FCs), the LLE has recently attracted renewed interest \cite{CS_KFC}. FCs consist of a set of equidistant spectral lines that can be used to measure light frequencies and hence time intervals more easily and precisely than ever before \cite{combs}. For this reason FCs open up a large variety of new applications ranging from optical clocks to astrophysics \cite{combs}. We explore the consequences of the present analysis for FC technology in a companion paper \cite{KFC_dark}.

As shown in Fig.~\ref{parameter_space}, in the normal dispersion regime rather large values of the detuning $\theta$ and pump power $\rho$ are required to obtain a sufficiently wide region of dark DSs (region IV) to observe such states experimentally. However, in recent years, the FC community has become increasingly successful at reaching the required values of pump power and detuning. As a result, dark DSs with different numbers of spatial oscillations (SOs) in their center (see, e.g., Fig.~\ref{LS4}) have been experimentally observed \cite{Xue_NP}. In Ref.\ \cite{Xue_NP} dark DSs were found using a normalized pump power $\rho \approx 2.5$ and normalized detuning $\theta \approx 5$. Figures \ref{parameter_space_H} and \ref{Hopfs} show that around these parameter values one can indeed find dark DSs with different numbers of SOs that can undergo oscillatory instabilities.

Our analysis provides a detailed map of the regions of existence and stability of dark DSs, which could serve as a guide for experimentalists to target particular DS solutions. We showed that dark DSs exist only in a well-defined zone within the wider region of bistability between two stable homogeneous solutions. Within this zone, dark DSs are organized in a bifurcation structure called a \textit{collapsed snaking} structure. The word "collapsed" refers to the fact that the region of existence of dark DSs shrinks exponentially with increasing number of SOs in the soliton profile (see Fig.~\ref{DarkS_4}). The collapse of the snaking structure implies that DSs with many SOs can only be found at the Maxwell point $\rho_M$, a fact that favors the observation of DSs with a single SO over that of broader DS with many SO. 

Although such a collapsed snaking structure persists for higher values of the detuning $\theta$, we also showed that narrow dark DSs with a low number
of SOs destabilize first as $\theta$ increases (Fig.~\ref{newfig8}) and start to oscillate in time. Therefore, at higher values of 
$\theta$ stable dark DSs found experimentally will most likely have an intermediate number of SOs. Our general analysis of the multistability of 
dark DSs can also explain previous numerical observations in Ref.\ \cite{Goday_chembo}, where it was shown that the pulse profile of dark
DSs becomes more distorted as the detuning increases. This could be due to the fact that stable dark DSs with a larger number of SOs are more 
likely to be found for higher values of the detuning.

\acknowledgments      
This research was supported by the Research Foundation--Flanders (FWO-Vlaanderen), by the Junior Mobility Programme (JuMo) of the KU Leuven, by the Belgian Science Policy Office (BelSPO) under Grant No. IAP 7-35, by the Research Council of the Vrije Universiteit Brussel, by the Spanish MINECO and FEDER under Grant Intense@Cosyp (FIS2012-30634), and by the National Science Foundation under grant DMS-1211953 (EK). We thank S. Coen and F. Leo for valuable discussions.

\section*{Appendix}

In this appendix we present details of the weakly nonlinear analysis near the RTB bifurcation at SN$_{hom,2}$ used to obtain analytically the spatially localized state in Eq.~(\ref{sech}). These states are solutions of the ODE system defined by
\begin{equation}\label{eq.1}
\bigg\{\begin{array}{c}
-\nu \displaystyle\frac{dV^2}{dx^2} -U+\theta V-V(U^2+V^2)+\rho=0 \\
\nu  \displaystyle\frac{dU^2}{dx^2}  -V-\theta U+U(U^2+V^2)=0.
\end{array}
\end{equation}

The bifurcation SN$_{hom,2}$ takes place at
\begin{equation}
 I_t=\displaystyle\frac{1}{3}(2\theta+\sqrt{\theta^2-3})
\end{equation}
and we consider a Taylor series expansion of $\rho$ around $I_t$:
\begin{equation}
 \rho(I_0)=\underbrace{\rho(I_t)}_{\rho_t}+\underbrace{\left(\frac{d\rho}{dI_0}\right)_{I_t}}_{=0}(I_0-I_t)+
 \underbrace{\displaystyle\frac{1}{2}\left(\frac{d^2\rho}{dI^2_0}\right)_{I_t}}_{\delta}\underbrace{(I_0-I_t)^2}_{\epsilon^2}+\cdots
\end{equation}
with 
\begin{equation}
 \rho_t=\sqrt{I_t^3-2\theta I_t^2+(1+\theta^2)I_t}.
\end{equation}
Because $\rho_t$ has a minimum at $I_t$, we have
$$\left(\displaystyle\frac{d\rho}{dI_0}\right)_{I_t}=0$$
$$\delta=\displaystyle\frac{1}{2}\left(\displaystyle\frac{d^2\rho}{dI^2_0}\right)_{I_t}=\displaystyle\frac{\sqrt{\theta^2-3}}{2\rho_t}>0.$$
We define a small parameter $\epsilon$ in terms of $\rho$,
\begin{equation}
 \epsilon=\sqrt{\frac{\rho-\rho_t}{\delta}},
\end{equation}
and use $\epsilon$ as an expansion parameter.

The localized states of interest can be written in the form
\begin{equation}\label{fullansatz}
 \left[\begin{array}{c}
U\\V
\end{array}\right]= \left[\begin{array}{c}
U\\V
\end{array}\right]^*+ \left[\begin{array}{c}
u\\v
\end{array}\right],
\end{equation}
with the spatially uniform states HSS given by
\begin{equation}\label{eq.HSS_up}
 \left[\begin{array}{c}
U\\V
\end{array}\right]^*= \left[\begin{array}{c}
U_t\\V_t
\end{array}\right]+ \epsilon\left[\begin{array}{c}
U_1\\V_1
\end{array}\right]+ \epsilon^2\left[\begin{array}{c}
U_2\\V_2
\end{array}\right]+...
\end{equation}
and the space-dependent terms by
\begin{equation}
 \left[\begin{array}{c}
u\\v
\end{array}\right]= \epsilon\left[\begin{array}{c}
u_1\\v_1
\end{array}\right]+ \epsilon^2\left[\begin{array}{c}
u_2\\v_2
\end{array}\right]+...
\end{equation}
We allow the fields $u_1$, $v_1$, $u_2$ and $v_2$ to depend on the slow variable $X\equiv\sqrt{\epsilon}x$. We first calculate the HSS terms and then do the same for the space-dependent terms.

\subsection*{Asymptotics for the uniform states}

Inserting the ansatz (\ref{eq.HSS_up}) in Eq.~(\ref{eq.1}), we obtain the correction to the HSS $A_0$ at any order in $\epsilon$. 

At order $\mathcal{O}(\epsilon^0)$ we obtain expressions for $U_t$ and $V_t$ as a function of $\theta$. At order $\mathcal{O}(\epsilon^1)$ we have
\begin{equation}\label{system.2_up}
L\left[\begin{array}{c}
U_1 \\ V_1\end{array}\right]=\left[\begin{array}{c}0\\ 0\end{array}\right],
\end{equation}
where 
\begin{equation}
L= \left[\begin{array}{cc}
0 & 0\\
-(\theta-I_t-2U_t^2)& -2
\end{array}\right]
\end{equation}
is a singular linear operator. Equation (\ref{system.2_up}) has an infinite number of solutions that can be written in the form
\begin{equation}\label{s}
\left[\begin{array}{c}
U_1 \\ V_1\end{array}\right]=\mu\left[\begin{array}{c}1\\ \eta\end{array}\right],
\end{equation}
where
\begin{equation}
\eta=-\displaystyle\frac{1}{2}(\theta-I_t-2U_t^2)
\end{equation}
and $\mu$ is obtained by solving the $\mathcal{O}(\epsilon^2)$ system.
At this order we obtain the equation
\begin{equation}
L\left[\begin{array}{c}
U_2 \\ V_2\end{array}\right] =
\left[\begin{array}{c}
2U_1V_1U_t +(2V_1^2+I_1)V_t-\delta\\
-(2U_1^2+I_1)U_t-2V_1U_1V_t
\end{array}\right],
\end{equation}\label{eq.2nd.orderHSS_up}
where $I_1\equiv U_1^2+V_1^2$. Because $L$ is singular, the previous equation has no solution unless a solvability condition is satisfied. This condition is given by 
\begin{equation}\label{mu_up}
\mu=\sqrt{\frac{\delta}{3\eta^2V_t+2\eta U_t+V_t}}.
\end{equation}

\subsection*{Asymptotics for the space-dependent states}

To calculate the space-dependent component of the weakly nonlinear state, we proceed in the same fashion. We 
insert the full ansatz for the asymptotic state, namely Eq.~(\ref{fullansatz}), into the system (\ref{eq.1}) and obtain, at order $\mathcal{O}(\epsilon^1)$,
\begin{equation}
 \underbrace{L\left[\begin{array}{c}
U_1 \\ V_1\end{array}\right]}_{=0}+ L\left[\begin{array}{c}
u_1 \\ v_1\end{array}\right]=\left[\begin{array}{c}
0 \\ 0\end{array}\right],
\end{equation}
where the first term on the LHS vanishes. The general solution of this equation is
\begin{equation}
 \left[\begin{array}{c}
u_1 \\ v_1\end{array}\right]=\left[\begin{array}{c}
U_1 \\ V_1\end{array}\right]\psi(X),
\end{equation}
with $\psi(X)$ a function to be determined at the next order.

\begin{figure}
\centering
\includegraphics[scale=0.9]{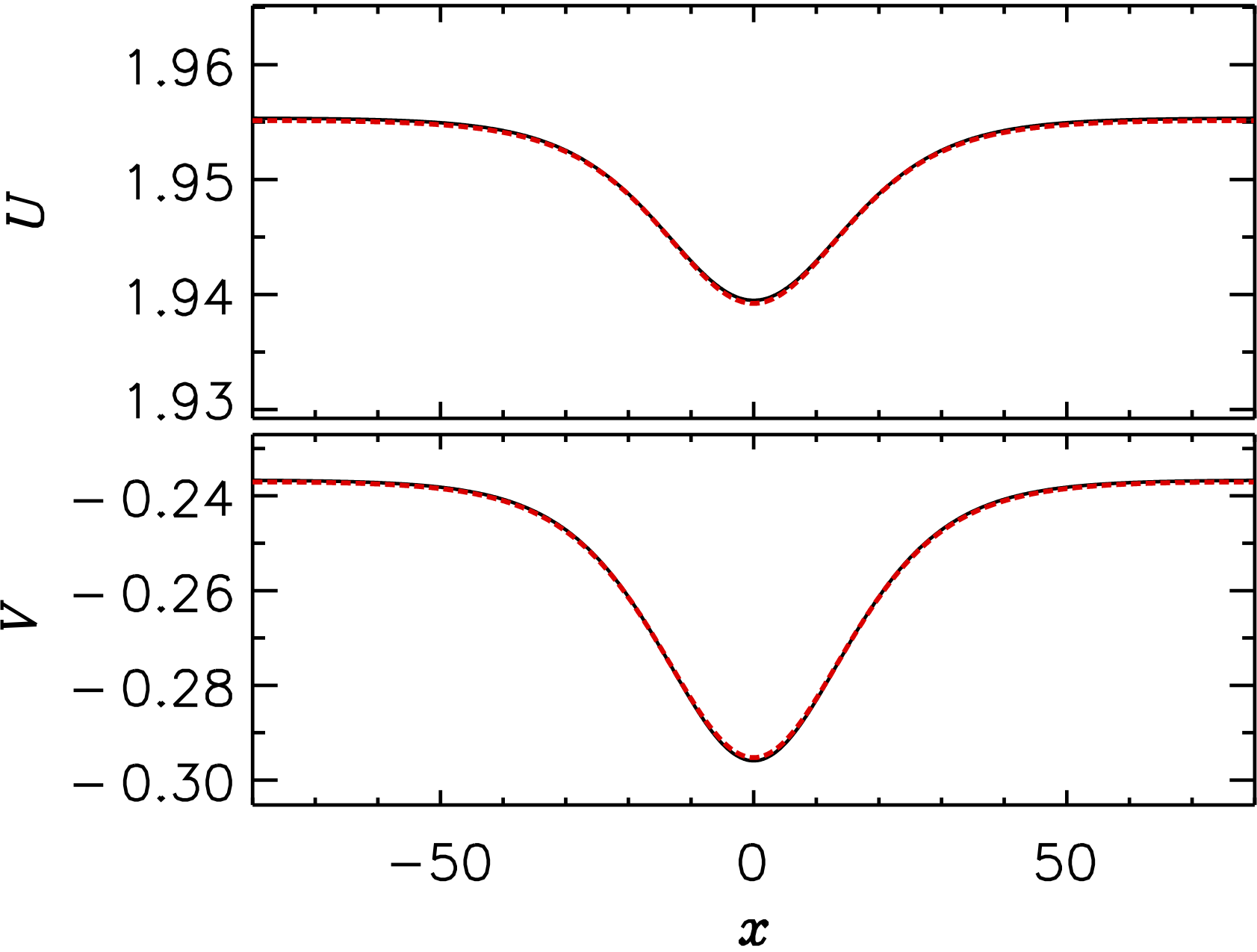}
\captionsetup{justification=raggedright,singlelinecheck=false}
\caption{(Color online) Asymptotic and exact hole solutions $A(x)\equiv U(x)+iV(x)$ close to SN$_{hom,2}$ for $\theta=4$ and $\rho=1.98388$. The black solid line shows the asymptotic solution for comparison with the numerically exact solution obtained by numerical continuation (red dashed line). The two lines are indistinguishable. }
\label{pulse_up}
\end{figure}

At order $\mathcal{O}(\epsilon^2)$
\begin{equation}\label{eq.2nd.spatial_up}
 L\left[\begin{array}{c}
u_2 \\ v_2\end{array}\right]=-\mathcal{P}_1\left[\begin{array}{c}
u_1 \\ v_1\end{array}\right]-\mathcal{P}_2\left[\begin{array}{c}
U_t \\ V_t\end{array}\right],
\end{equation}
with the linear operators
\begin{equation}
 \mathcal{P}_1=\left[\begin{array}{cc}
-(2U_tV_1+2U_1V_t) & -(\nu\partial_X^2+6V_tV_1+2U_tU_1)\\
\nu\partial_X^2+6U_tU_1+2V_tV_1& 2V_tU_1+2U_tV_1
\end{array}\right]
\end{equation}
and
\begin{equation}
 \mathcal{P}_2=\left[\begin{array}{cc}
-2v_1u_1 & -(3v_1^2+u_1^2)\\
3u_1^2+v_1^2 & 2v_1u_1
\end{array}\right].
\end{equation}
Because $L$ is singular, Eq.~(\ref{eq.2nd.spatial_up}) has no solution unless another solvability condition is satisfied. In the present case, this condition reads 
\begin{equation}\label{solvability.2_up}
 \left[\begin{array}{cc}
1 & 0\end{array}\right]\mathcal{P}_1\left[\begin{array}{c}
u_1 \\ v_1\end{array}\right]+ \left[\begin{array}{cc}
1 & 0\end{array}\right]\mathcal{P}_2\left[\begin{array}{c}
U_t \\ V_t\end{array}\right]=0.
\end{equation}

After some algebra, Eq.~(\ref{solvability.2_up}) reduces to an ordinary differential equation for $\psi(X)$,
\begin{equation}\label{psi.eq.1_up}
\alpha_1\psi''(X)+\alpha_2\psi(X)+\alpha_3\psi^2(X)=0,
\end{equation}
where
\begin{equation}
\begin{array}{ccc}
\alpha_1=-\nu V_1, & \alpha_2=-2\delta, & \alpha_3=-\delta.
\end{array}
\end{equation}
This equation has solutions homoclinic to $\psi=0$ given by
\begin{equation}
 \psi(X)=-3\textnormal{sech}^2\left(\displaystyle\frac{1}{2}\sqrt{-\frac{\alpha_2}{\alpha_1}}(X-X_0)\right),
\end{equation}
representing a hole in the spatially uniform state located at $X=X_0$, hereafter at $X=0$. Since $X\equiv\sqrt{\epsilon}x$ and $\epsilon\equiv\sqrt{\displaystyle\frac{\rho-\rho_t}{\delta}}$ the corresponding first order spatial correction is given by
\begin{equation}
\left[\begin{array}{c}
u_1\\v_1
\end{array}\right]= 
-3\mu\left[\begin{array}{c}
1\\ \eta\end{array}\right]{\textnormal{sech}}^2
\left[ \displaystyle\frac{1}{2}\sqrt{-\frac{\alpha_2}{\alpha_1}}\left(\displaystyle\frac{\rho-\rho_{t}}{\delta}\right)^{1/4}x\right].
\end{equation}
The resulting asymptotic solution for $\theta=4$ and $\rho=1.98388$ is shown in Fig.~\ref{pulse_up} (black solid lines). The corresponding numerically exact solution, obtained using numerical continuation, is shown in red dashed lines. The agreement is excellent.

For $\sqrt{3}<\theta<2$ the saddle node SN$_{hom,1}$ is also a RTB bifurcation and the same asymptotic calculation can therefore be used to compute the DSs present near this bifurcation. A related calculation can be used to compute the DS profiles near the point HH \cite{BuYoKn}.

\end{document}